\documentclass[twocolumn,twocolappendix]{aastex63}

\usepackage{graphicx}	
\usepackage{color}
\usepackage{amsmath}	
\usepackage{amssymb}	
\usepackage{mathrsfs}
\usepackage{mathrsfs}

\newcommand{\mbh}{M_{\bullet}}
\newcommand{\msun}{M_\odot}
\newcommand{\mh}{M_{\rm h}}
\newcommand{\mstar}{M_\star}

\newcommand{\mdot}{\dot{M}_{\bullet}}
\newcommand{\msunyr}{M_\odot~{\rm yr}^{-1}}
\newcommand{\mdote}{\dot{M}_{\rm Edd}}

\newcommand{\mpc}{{\rm Mpc}}

\newcommand{\ergs}{{\rm erg~s}^{-1}}

\newcommand{\D}{{\rm d}}

\newcommand{\Muv}{M_{\rm 1450}}

\newcommand{\K}{{\rm K}}
\newcommand{\beq}{\begin{equation}}
\newcommand{\eeq}{\end{equation}}

\newcommand{\fobsc}{f_{\rm obsc}}
\newcommand{\fb}{f_{\rm b}}
\newcommand{\fstar}{f_{\star}}
\newcommand{\fseed}{f_{\rm seed}}

\defcitealias{Harikane_2023}{H23}

\shorttitle{Growth of High Redshift Quasars}
\shortauthors{Li et al.}

\begin{document}

\title{Reconstruction of Cosmic Black Hole Growth and Mass Distribution from Quasar Luminosity Functions at $z>4$:
Implications for Faint and Low-mass Populations in JWST}

\correspondingauthor{Wenxiu Li, Kohei Inayoshi}
\email{wenxiuli@pku.edu.cn, inayoshi@pku.edu.cn}

\author[0000-0002-1044-4081]{Wenxiu Li}
\affiliation{Kavli Institute for Astronomy and Astrophysics, Peking University, Beijing 100871, China}
\affiliation{Department of Astronomy, School of Physics, Peking University, Beijing 100871, China}

\author[0000-0001-9840-4959]{Kohei Inayoshi}
\affiliation{Kavli Institute for Astronomy and Astrophysics, Peking University, Beijing 100871, China}

\author[0000-0003-2984-6803]{Masafusa Onoue}
\altaffiliation{Kavli Astrophysics Fellow}
\affiliation{Kavli Institute for Astronomy and Astrophysics, Peking University, Beijing 100871, China}
\affiliation{Kavli Institute for the Physics and Mathematics of the Universe (Kavli IPMU, WPI), The University of Tokyo, Chiba 277-8583, Japan}

\author[0000-0001-7759-6410]{Wanqiu He}
\affiliation{National Astronomical Observatory of Japan, Mitaka, Tokyo 181-8588, Japan}

\author[0000-0001-5063-0340]{Yoshiki Matsuoka}
\affiliation{Research Center for Space and Cosmic Evolution, Ehime University, Matsuyama, Ehime 790-8577, Japan}

\author[0000-0003-0230-6436]{Zhiwei Pan}
\affiliation{Kavli Institute for Astronomy and Astrophysics, Peking University, Beijing 100871, China}
\affiliation{Department of Astronomy, School of Physics, Peking University, Beijing 100871, China}

\author[0000-0002-2651-1701]{Masayuki Akiyama}
\affiliation{Astronomical Institute, Tohoku University, Sendai, 980-8578, Japan}

\author[0000-0001-9452-0813]{Takuma Izumi}
\affiliation{National Astronomical Observatory of Japan, Mitaka, Tokyo 181-8588, Japan}
\affiliation{Astronomical Science Program, The Graduate University for Advanced Studies, SOKENDAI, 2-21-1 Osawa, Mitaka, Tokyo 181-8588, Japan}
\author[0000-0002-7402-5441]{Tohru Nagao}
\affiliation{Research Center for Space and Cosmic Evolution, Ehime University, Matsuyama, Ehime 790-8577, Japan}

\begin{abstract}
The evolution of the quasar luminosity function (QLF) is fundamental to understanding the cosmic evolution of black holes (BHs) 
through their accretion phases.
In the era of the James Webb Space Telescope (JWST), Euclid, and Nancy Grace Roman Space Telescope, their unprecedented detection 
sensitivity and wide survey area can unveil the low-luminosity quasar and low-mass BH population,
and provide new insights into quasar host galaxies.
We present a theoretical model describing BH growth from initial seeding at $z\gtrsim 20$ to $z\sim 4$,
incorporating the duration of accretion episodes, the distribution of Eddington ratios, 
and the mass dependency of BH accretion rates.
By constraining the model parameters with the observed QLFs at $4\leq z\leq 6$ across a wide UV luminosity range, 
we find that the high-redshift BH population grows rapidly at $z\gtrsim 6$, and decelerates the pace in subsequent epochs.
Toward lower redshifts ($z<6$), mass-dependent accretion inhibits the growth of high-mass BHs with $\mbh>10^8~\msun$,
leading to mass saturation at $\mbh\gtrsim 10^{10}~\msun$.
We predict the BH mass function down to $\mbh \sim 10^6~\msun$ for both unobscured and obscured quasar populations at $4\leq z \leq 11$,
offering a benchmark for future observational tests.
Our model accounts for the presence of both bright and faint quasars at $z>4$, including those discovered by JWST.
Furthermore, our findings suggest two distinct pathways for the early assembly of the BH-galaxy mass correlation:
the population with a BH-to-stellar mass ratio near the local value of $\mbh/\mstar \simeq 5\times 10^{-3}$ maintains a proximity to the relation via moderate growth, 
while the population that begins to grow above the local relation becomes as overmassive as $\mbh/\mstar \sim 0.01-0.1$ by $z\sim 6$ via rapid mass accretion.
\end{abstract}


\section{Introduction}
The quasar luminosity function (QLF) at high redshifts encodes key information on the radiative properties of black hole (BH) growth led by mass accretion.
Our current understanding of unobscured quasars at $z\gtrsim 4$ significantly benefits from wide-field surveys such as the Sloan Digital Sky Survey \citep[SDSS;][]{Jiang_2016, Wu_Shen_2022}, the Pan-Starrs 1 \citep{Banados_2016, Banados_2023}, and the Dark Energy Spectroscopic Instrument \citep{Yang_2023}. 
Among these efforts, the Hyper Suprime-Cam Subaru Strategic Program (HSC-SSP; \citealt{HSCSSP_2018}) has the unique advantage in finding high-$z$ low-luminosity quasars.
In combination with the SDSS, the HSC-SSP has constrained $4\lesssim z\lesssim 7$ QLF down to the faint end 
($\Muv \simeq -22$ mag in absolute UV magnitude), thanks to the deep sensitivity of the 8.2m Subaru telescope and its wide ($\sim1,100$ deg$^2$) coverage (\citealt{Akiyama_2018} for $z\sim4$, \citealt{Niida_2020} for $z\sim5$, \citealt{Matsuoka_2018} for $z\sim6$, and \citealt{Matsuoka_2023} for $z\sim7$).
These studies have provided the constraints on the quasar contribution to the cosmic X-ray and infrared background \citep[e.g.,][]{Hauser_Dwek_2001,Shen_2020}, 
and suggested that the ionizing photon budget from those quasars is not sufficient to complete cosmic reionization
due to a rapid decay of the quasar abundance at $z\geq 6$ (e.g., \citealt{Wang_2019}, \citealt{Kim_2022}, and \citealt{Schindler_2023}, but also an alternative interpretation by 
\citealt{Fontanot_2023}).

The James Webb Space Telescope (JWST) has been revolutionizing our understanding of the nature of early galaxies and 
their co-evolution with massive BHs in the nuclei.
The unprecedented sensitivity of JWST has not only allowed for the detection of ultra high-redshift galaxies at $z>10$
\citep[e.g.,][]{Adams_2022, Castellano_2022, Donnan_2022, Finkelstein_2022, Harikane_2022c, Naidu_2022,Fujimoto_2023,Arrabal_Haro2023,Harikane_2023_galaxy},
but also enabled the discovery of low-luminosity active galactic nuclei (AGNs) at $z>5$, which were hidden in the pre-JWST era. 
Spectroscopic follow-up observations have provided estimates of 
the nuclear BH masses for these sources, including the discovery of the least-massive known BH with $\mbh \sim 10^7~\msun$ 
at the end of cosmic reionization \citep{Onoue_2023,Kocevski_2023} and the most distant quasar at $z\simeq 8.6$ \citep{Larson_2023}.
\citet{Maiolino_2023} recently reported the NIRSpec observation of GN-z11, an exceptionally luminous galaxy at $z=10.6$, 
revealing the high ionization [Ne~{\sc iv}] $\lambda$2423 transition and semi-forbidden nebular lines tracing clouds of 
broad-line regions of an AGN. The Mg~{\sc ii}-based virial BH mass is estimated as $\mbh \simeq 1.6\times 10^6~\msun$.
Moreover, Matthee et al. (2023) discovered twenty broad-line AGNs via slitless spectroscopic observations in the EIGER and FRESCO fieds,
and \citet{Maiolino_2023Jades} found a sample of twelve broad-line AGNs in the JADES survey.
The BH masses infered from the broad Balmer lines range from $10^6~\msun$ to $10^7 ~\msun$.
The findings of low-luminosity AGNs and low-mass BHs put constraints on the mass distribution for seeds of high-$z$ quasars 
and the early stage of BH-galaxy co-evolution \citep[e.g.,][]{Inayoshi_ARAA_2020,Volonteri_2021,Trinca_2023,Schneider_2023}.
Intriguingly, the number density of broad-line faint AGNs detected by JWST is higher than extrapolation of the QLF 
constructed by HSC-SSP \citep{Kocevski_2023,Matthee_2023}, and is comparable to the X-ray selected AGNs \citep{Giallongo2019}.
The abundance is estimated to be even higher in \citet{Maiolino_2023Jades} and \citet{Harikane_2023}.
These abundance estimates indicate a large population of faint AGNs buried in their host galaxies.

On the other hand, the shape and evolution of the BH mass function (BHMF) remain inadequately constrained, particularly at high redshifts of $z \gtrsim 2$, 
where observation data have been focused on the high-mass end of $\mbh > 10^8~\msun$ 
\citep{Vestergaard_2008, Willott_2010,Kelly_Shen_2013,Wu_2022}.
This limitation arises from the fact that dedicated spectroscopic observations for a large sample are required to 
measure the virial BH masses through Balmer lines or metal lines such as Mg~{\sc ii} and C~{\sc iv},
thereby making the construction of the BHMF observationally expensive.
Additionally, correction for incompleteness in the BHMF at low mass ends is more challenging than that for the QLF at faint ends,
as the BH mass does not directly correlate with the source brightness.
Nevertheless, extensive efforts have been made to identify BHs with masses of $\mbh \lesssim 10^8~\msun$ at $z>4$ from the ground 
\citep[e.g.,][]{Kim_2018, Onoue_2019} and recently by the JWST \citep{Kocevski_2023, Harikane_2023}.
\citet{He_2023} reconstructed the $z\sim4$ BHMF based on their HSC-SSP sample, pushing the mass boundary down to $\mbh \sim 10^{7.5}~\msun$.
The cosmic evolution of the QLF and BHMF provides the key to the origin of SMBHs and their growth mechanisms, in comparison 
to theoretical models of the low-mass end of the BHMF \citep[e.g.,][]{Ricarte_2018,Basu_2019,Kim_Im2021,Li_2023,Ni2022,Trinca_2022}.
Furthermore, the ongoing JWST observations have pushed the mass boundary even down to $\mbh\simeq 10^6$ -- $10^7~\msun$ at $z\gtrsim 4-6$, 
potentially uncovering even fainter and more representative quasars \citep[e.g.,][]{Larson_2023,Maiolino_2023,Maiolino_2023Jades,Matthee_2023}. 

A pioneer work by \cite{Small_Blandford_1992} studied the cosmic evolution of the BHMF using the observed QLF as a flux term of the continuity equation.
The continuity equation approach has been widely used to determine the buildup of cosmic BHs,
probing testbeds for the distributions of the radiative efficiency, Eddington ratio, and duty cycle
\citep{Yu_Tremaine_2002,Marconi_2004,Yu_Lu_2004,Shankar_2010,Shankar_2013,Aversa_2015,Ricci_2017emissivity,Ananna_2020}.
On the other hand, empirical models attempt to follow the evolution of dark matter halos, galaxies, and BHs self-consistently,
and set the constraints on the underlying BH abundance and their growth properties \citep{Behroozi_Silk_2018,Zhang_2023}.

In this paper, we present a theoretical model that describes the growth of BHs through mass accretion,
spanning from initial seeding epochs at $z\gtrsim 20$ to $z\simeq 4$.
Utilizing the Markov Chain Monte Carlo (MCMC) fitting technique, we optimize the model parameters for BH mass accretion
to align with the observed QLFs at $4\leq z \leq 6$ across a wide UV luminosity range ($-29< \Muv <-24$).
The best-fitted model successfully reproduces the cosmic evolution of QLFs observed at those redshifts and  
suggests rapid growth of BH populations at $z>6$ followed by a deceleration in subsequent epochs.
Using the result, we reconstruct the BHMF down to $M_\bullet \simeq 10^6~\msun$ for unobscured and obscured quasar populations at $4\leq z \leq 11$
and explore the assembly of cosmic BH mass density by integrating the BHMF.
Moreover, our seeding and growth model offers the formation pathways of BHs in two distinct populations: 
these observed in the brightest quasars at $z>6$ and the faint quasars identified at $z>4$ through JWST observations.
The BH growth model, calibrated with the most updated QLF studies, sheds light on the early assembly of the BH-to-galaxy mass correlation 
and provides insights into the evolutionary trajectories of both undermassive and overmassive BH populations relative to the local 
empirical correlation.

This paper is organized as follows. 
In Section~\ref{sec:method}, we describe the procedure to constrain the BH growth model by the observed QLFs.
In Section~\ref{sec:result}, we present the fitting result of the QLF and show the predicted shape of the BHMF 
at $4\lesssim z \lesssim 11$ in the mass range of $\mbh \gtrsim 10^6~\msun$,
as well as the redshift dependence of the cosmic BH mass density.
In Section~\ref{sec:discussion}, we illustrate the evolutionary tracks of individual BHs based on the best-fit growth model,
and discuss the mass assembly of high-$z$ bright and faint quasars and the establishment of their BH-to-stellar mass ratio.
Finally, we summarize our findings in Section~\ref{sec:summary}.
Throughout this paper, we adopt the cosmological parameters from \cite{Planck2016},
i.e., $\Omega_{\mathrm{m}}=0.307,~\Omega_{\Lambda}=0.693,~
\Omega_{\mathrm{b}}=0.0486$, and $H_0=67.7 \mathrm{~km} \mathrm{~s}^{-1} \mathrm{Mpc}^{-1}$.
All magnitudes quoted in this work are in the AB system.

\section{Method}
\label{sec:method}
In this study, we build upon the work of \cite{Li_2023} to explore the QLF and BHMF from early seeding epochs at $z>20$ down to $z\sim 4$.
Our findings, especially on the BH population at the faint and low-mass end at high redshifts, provide crucial insights into seed BHs 
and set a benchmark for observational studies of BH evolution.
Our method adopted from the previous work consists of three main steps, detailed in the following sections:
the initial conditions for the BHMF (Section~\ref{sec:method_initial}), 
the evolution of the BHMF by mass accretion (Section~\ref{sec:method_BH}), and
the MCMC parameter optimization with the observed QLFs and BHMFs (Section~\ref{sec:method_fitting}).

\subsection{Initial conditions for the BHMF}
\label{sec:method_initial}

Our model begins with the BHMF at $z=20$ calculated by \citet{Li_2023}.
We concentrate on BH seeds forming in progenitor dark matter (DM) halos, which eventually grow to be high-$z$ quasar host galaxies.
These galaxies have halo masses ranging from $M_{\rm h}\simeq  10^{11}~\msun$ to $10^{13}~\msun$ at $z=6$.
For each of these parent halos, we generate merger trees backward in time using the {\tt GALFORM} semianalytic algorithm 
based on the extended Press-Schechter formalism \citep{Parkinson_2008}.

Within each tree, we initiate primordial gas clouds and simulate the dynamical, thermal, and chemical evolution \citep{Li_2021}.
The model focuses on overdense regions of the universe, where those progenitor halos are exposed to intense H$_2$-photodissociating 
radiation from nearby star-forming galaxies and heat the interior gas by successive mergers and 
baryonic streaming motion relative DM (see also \citealt{Lupi_2021}). 
Those effects suppress H$_2$ formation and cooling, preventing the gas cloud from collapsing into stars and thereby increasing the cloud Jeans mass.
When a cloud begins to collapse under these circumstances, stars form and grow rapidly through mass accretion at rates of 
$\sim 1-10~\msunyr$ \citep{Regan_2014, IOT_2014}, eventually leaving behind massive BHs \citep{Hosokawa_2013, Toyouchi_2023}.
\cite{Li_2023} analyzed the individual evolution and seeding mechanisms of these BHs, and 
found that the mass distribution function is influenced by the diversity in parent halo properties and environmental factors.
As a result, the BHMF for these seed populations is well approximated with a Salpeter-like distribution and 
extends the upper mass to $M_\bullet \gtrsim 10^5~\msun$.

In the progenitor halos of quasar hosts, the formation of seed BHs ceases by $z\sim 20$.
Beyond this point, the comoving number density of seed BHs remains constant at $\simeq 10^{-3}~{\rm Mpc}^{-3}$,
assuming no BH mergers occur.
\citet{Li_2023} explored the growth process of these BHs, considering that a fraction $\fseed (\leq 1)$ of the BHs contribute 
to the assembly of SMBHs at later epochs; namely,
\begin{equation}
    N_{\rm seed} = 10^{-3}\left(\frac{\fseed}{1.0}\right) ~ {\rm Mpc}^{-3}.  
\end{equation}
We examine two cases with $\fseed=0.1$ and $1.0$ in this paper, given the current theoretical and observational uncertainties.
This parameter not only influences the BH growth dynamics but also exhibits a correlation with other model parameters.
Intriguingly, higher values of $\fseed$ lead to a greater abundance of quasars at the faint end ($\Muv > -22$), 
aligning with the numerous faint AGNs observed in recent JWST observations \citep{Kocevski_2023, Harikane_2023}. 
See more details in Section~\ref{sec:result}.

\subsection{Construction and evolution of the BHMF}
\label{sec:method_BH}
We characterize the growth of each BH by a minimal number of free parameters, giving the accretion rate by
\begin{equation}
  \label{eq:mdot}
  \mdot = \lambda f(\mbh) \mdote,
\end{equation}
where $\lambda$ is the luminosity-based Eddington ratio, and $\mdote \equiv L_{\rm Edd}/\eta_0 c^2$ 
is the Eddington accretion rate with a radiative efficiency of $\eta_0=0.1$ \citep{Shakura_Sunyaev_1973}.
To model a non-exponential growth manner, we introduce a functional form: 
\begin{equation}
    \label{eq:f_M}
    f(\mbh) = \frac{2}{1+\left(\mbh /M_{\rm \bullet,c} \right)^\delta}, 
\end{equation}
where $M_{\rm \bullet,c}=10^8~\msun$ is adopted \citep[e.g.,][]{Ueda_2014}.
With a positive value of $\delta$, the BH growth at the high mass end is suppressed,
while the growth speed for less massive BHs is accelerated.
In the limit of $\delta \ll 1$, where $f(\mbh) \simeq 1$ is nearly independent of $M_{\rm \bullet,c}$,
the model in Eq.~(\ref{eq:mdot}) reduces to an exponential growth with an $e$-folding timescale of
$t_{\rm S} =  \mbh/\mdote \approx 45$ Myr.

Quasar activity is thought to take place episodically with accretion bursts triggered by gas inflows to the galactic nuclei
and subsequent gas consumption \citep{Di_Matteo_2005, Hopkins_2005, Hopkins_Quataert_2010}.
The episodic behavior of individual quasar light curves is directly reflected in the diversity seen
in the Eddington ratio distribution function (ERDF) of observed quasars \citep{Hopkins_2005_LF}.
In this work, we assume the ERDF to be characterized with a Schechter function in a range of $\lambda_{\rm min}\leq \lambda < \infty$,
\begin{equation}
  \label{eq:Pl}
  \frac{\D P}{\D \ln \lambda} \propto
  \left(\frac{\lambda} {\lambda_0} \right)^\alpha \exp{\left(-\frac{\lambda}{\lambda_0}\right)},
\end{equation}
where $\lambda_0$ and $\alpha$ are considered to be free parameters in the following analysis.
The minimum Eddington ratio is set to $\lambda_\mathrm{min}=0.01$, motivated from simulations of quasar activity \citep{Navak_2011},
showing a luminosity decline towards $\lambda \sim 0.1-0.01$ within $\simeq 100$ Myr post-peak activity.
However, this choice is tentative and leads to uncertainties in the estimated fraction of inactive BHs.
We examine the scenario where $\alpha=-0.12$ (the best-fit value at $z\sim 6-5$), finding that populations with lower $\lambda$ 
contribute significantly to the total BH abundance.
For instance, setting $\lambda_\mathrm{min}=10^{-3}$ results in about 50\% of the total BHs being classified as inactive ($\lambda < 0.01$),
which is a twofold reduction in the active BH count compared to the baseline scenario of $\lambda_\mathrm{min}=10^{-2}$.
We calculate the frequency of mass accretion bursts for individual BHs by assigning a duration $\tau$
to each burst, during which the BH accretes mass at a constant Eddington ratio $\lambda$, following the ERDF in Eq.~(\ref{eq:Pl}).
This approach allows us to model varying BH growth rates over different $\tau$ periods, capturing the episodic nature of accretion bursts. 
This method simplifies the process by sidestepping the complex modeling of galaxy assembly, gas feeding, and BH feedback in quasar progenitor halos.

To model the time evolution of the BH mass function, we focus on the growth of existing BHs via mass accretion.
The distribution function $\D \Phi/\D M_{\bullet, 0}$ at $t=t_0$ evolves to $t=t_0+\tau$ following the equation:
\begin{equation}
  \label{eq:dpdm}
  \frac{\D\Phi}{\D \mbh} = \int
   \frac{\D P}{\D \ln \lambda}\Big|_{\lambda_\ast} \cdot
   \frac{\D \ln \lambda_\ast}{\D  \mbh}\Big|_{\tau, M_{\bullet,0}} \cdot
  \frac{\D \Phi}{\D M_{\bullet, 0}}~\D M_{\bullet, 0},
\end{equation}
where $\lambda_\ast$ is the Eddington ratio required for a BH with $M_{\bullet,0}$ to grow to $\mbh$ in $\tau$,
calculated from Eqs.~(\ref{eq:mdot}) and (\ref{eq:f_M}).
This integration method excludes the effects of newly-born BHs and BH coalescence,
and is thus carried out without a source or sink term ensuring BH number density conservation.
The BHMF evolution is calculated with multiple accretion episodes, each of which lasts a time duration of $\tau$ and resamples $\lambda$ 
following Eq.~(\ref{eq:f_M}), down to the observed redshift.
For the QLF generation for unobscured populations, we convolve the BHMF with the quasar ERDF from Eq.~(\ref{eq:Pl}) and 
take into account the obscured fraction $\fobsc$.
The fraction $\fobsc$, based on X-ray observations up to $z\sim 5$ by \citet{Ueda_2014}, is incorporated
along with a conversion from hard X-ray to bolometric luminosity \citep{Duras2020}. 
The rest-frame ultraviolet (UV) absolute magnitude is calculated as 
\begin{equation}
  \label{eq:M1450}
  \Muv= -21.0-2.5 \log  \left(\frac{L_{\rm bol}}{10^{45}~\mathrm{erg~s}^{-1}} \right) ~[\rm{mag}],
\end{equation}
where $L_{\rm bol}=\lambda L_{\rm Edd}$ is the bolometric quasar luminosity. 
The unobscured QLF is derived as
\begin{equation}
\label{eq:dn_dM1450}
\Phi_{\Muv} 
 = (1 -\fobsc)  
\int \frac{\D P}{\D \ln \lambda}\Big|_{\tilde{\lambda}}  \cdot
\frac{\D \ln \tilde{\lambda}}{\D \Muv} \cdot
 \Phi_{\mbh} \D \log \mbh,
\end{equation}
where the values of $\tilde{\lambda}=\lambda(\Muv, \mbh)$ and $\D \ln \tilde{\lambda}/\D \Muv$ are calculated analytically 
from Eq.~(\ref{eq:M1450}).

In our model, the parameter $\tau$ plays a crucial role in shaping the QLF and BHMF,
as it controls the average BH growth rate by recurring samplings of the Eddington ratio $\lambda$.
A smaller $\tau$ leads to more frequent changes in $\lambda$, and thus the average growth rate is closer to the peak of 
the ERDF, influenced by $\lambda_0$ and $\alpha$.
On the other hand, a larger $\tau$ allows for longer periods of high-rate BH growth at a non-negligible likelihood.
This mechanism leads to a modulation of the BHMF and QLF, where smaller (larger) values of $\tau$ make
the slopes of the high mass and bright end steeper (shallower).
This model with variable Eddington ratios allows BHs to undergo episodic growth with both high and low accretion rates,
unlike approaches in \citet{Shankar_2010} that assign a single $\lambda$ and constrain the model by the QLF and clustering \citep{Haiman_Hui_2001}.
Our method inherently accounts for inactive BHs with luminosities below the detection limits of quasar surveys, 
when correlating QLFs with BHMFs.

\subsection{Observational data and fitting analysis}
\label{sec:method_fitting}

\begin{deluxetable*}{lcccccc}
\tablecolumns{1}
\renewcommand\thetable{1} 
\tablewidth{0pt}
\tablecaption{Best-fit parameters for BH growth\label{tab:para}}
\tablehead{
  \colhead{$\fseed$} & \colhead{Redshift} &   \colhead{$\tau$ (Myr)}  & \colhead{$\log\delta$} &  \colhead{$\lambda_0$}  & \colhead{$\alpha$}  
}
\startdata
0.1 & $4\leq z < 5$    &  151 ($111^{+50}_{-67}$)  &   $-0.26$ ($-0.26^{+0.09}_{-0.09}$)    &   0.42  ($0.45^{+0.17}_{-0.13}$)   &    $-0.02$ ($0.00^{+0.25}_{-0.20}$)   \\
& $5\leq z < 6$    &  195 ($190^{+49}_{-70}$)  &   $-0.23$ ($-0.22^{+0.10}_{-0.10}$)    &   0.50 ($0.58^{+0.25}_{-0.16}$)    &    $-0.12$ ($-0.18^{+0.19}_{-0.17}$)   \\
& $6 \leq z$       &  20.1 ($27.0^{+19.4}_{-9.8}$)&   $-2.98$ ($-2.36^{+0.65}_{-0.55}$)    &   0.89 ($0.80^{+0.28}_{-0.23}$)    &    0.12    ($0.11^{+0.17}_{-0.14}$)   \vspace{1mm}\\
& $4\leq z < 6^\dag$    &  142 ($137^{+33}_{-40}$)  &   $-0.24$ ($-0.23^{+0.07}_{-0.08}$)    &   0.53  ($0.58^{+0.16}_{-0.12}$)   &    $-0.12$ ($-0.17^{+0.12}_{-0.11}$)   \\
\hline
 1 & $4\leq z < 5$    &  362 ($323^{+48}_{-63}$)  &   $-0.29$ ($-0.26^{+0.10}_{-0.08}$)    &   0.28  ($0.33^{+0.19}_{-0.09}$)   &    $-0.30$ ($-0.34^{+0.23}_{-0.20}$)   \\
   & $5\leq z < 6$    &  200 ($193^{+78}_{-52}$)  &   $-0.39$ ($-0.38^{+0.09}_{-0.11}$)    &   0.44  ($0.51^{+0.24}_{-0.14}$)   &    $-0.41$ ($-0.48^{+0.22}_{-0.18}$)   \\
 & $6 \leq z$    &  23.1 ($29.9^{+21.0}_{-10.2}$) &   $-2.97$ ($-2.50^{+0.55}_{-0.43}$)    &   0.95  ($0.85^{+0.32}_{-0.28}$)   &    $-0.06$ ($-0.07^{+0.16}_{-0.12}$)  \\
\enddata
\tablecomments{
The best-fit parameters optimized by MCMC sampling for the three redshift ranges, for the cases with $\fseed=0.1$ and 1.0.
The 16\%, 50\%, and 84\% quantiles in the cumulative distribution are shown in the parentheses.
$\dag$ The result for the case where only the QLF data at $z=6$ and $z=4$ are taken into account for the parameter fitting with the case of $\fseed=0.1$.}
\end{deluxetable*}

We calibrate the four parameters in our model ($\alpha$, $\lambda_0$, $\tau$, and $\delta$) by comparing the model-generated
QLF and BHMF with obsrvational data at $z\gtrsim 4$.
We use the observed QLF data at $z\sim 4$ from \citet{Akiyama_2018}, at $z \sim 5$ from \citet{Niida_2020}, 
and $z \sim 6$ from \citet{Matsuoka_2018}, covering a wide UV magnitude range of $-30<\Muv < -22$.
These datasets, with their specific magnitude bins and error estimates, provide robust constraints for our BH growth model.
To reduce potential biases and selection effects in those quasar surveys (e.g., point source against extended source),
our fitting is limited to UV magnitudes brighter than $\Muv<-24$.
Additionally, we incorporate the BHMF data at $z\sim 6$ from \citet{Willott_2010} to refine our model for BH masses 
at $10^8~\msun < \mbh < 10^{10}~\msun$, thereby improving parameter degeneracy \citep{Li_2023}.

It is worth noting that the redshifts for the quasar samples used in constructing the binned QLFs do not exactly match the integer values ($z=6$, $5$, and $4$).
Instead, these samples have mean redshifts of $z=$ 6.1 \citep{Matsuoka_2018}, 
4.9 \citep{Niida_2020}, and 3.9 \citep{Akiyama_2018}, respectively.
The completeness function for quasar selection is dependent on both redshift and luminosity, which implies 
that correcting for the small redshift variations in each luminosity bin (corresponding to an error of $\Delta t \lesssim 30$ Myr) 
could introduce systematic errors.
However, our fitting results based on these mean redshifts align closely with those using integer redshifts,
falling within the statistical errors of the binned QLF.

In the MCMC fitting, the posterior probability is evaluated by using a $\chi^2$-type value defined as 
\beq
  \chi^2 = \sum_{i}
  \frac{\left(\log{\Phi^{\rm mod}_{i}} - \log{\Phi^{\rm obs}_{i}}\right)^2}{(\log{\Phi^{\rm err}_{i}})^2}
  - 2 \sum_{a} \ln P^{\rm prior}_a.
\eeq
This formula includes the classical $\chi^2$ value for dispersion between observed QLF data and model predictions, 
and a term for parameter deviation from prior distributions.
We set uniform priors for $10 \leq \tau/({\rm Myr}) \leq 400$ and $-4\leq \log \delta \leq 0$, and adopt 
prior distributions for $\lambda_0$ and $\alpha$ as in \citet{Li_2023}.
The optimal parameters are identified at the minimum $\chi^2$ value, indicating the highest probability of reproducing 
the observed QLF data.
Our analysis involves fitting BH growth parameters for two distinct periods:
from $z=6$ to $z=5$ (for a time duration of $\Delta t_{\rm H}\simeq 237$ Myr), and 
from $z=5$ to $z=4$ (for $\Delta t_{\rm H} \simeq 361$ Myr).
This treatment minimizes discrepancies between modeled and observed QLFs at these redshifts. 
The best-fit parameters for each period differ, reflecting variation of the BH growth patterns.

Additionally, we examine the fit of the BH growth model using only the QLF data at $z=4$, while excluding the $z=5$ QLF.
In this case, the modeled QLF at $z= 5$ shows a significant underestimate of the quasar abundance compared to the observed data.
This result suggests that the quasar growth at $5<z<6$ is more rapid than the average rate at $4<z<6$,
indicating that a uniform growth pattern across these redshifts is not plausible.
Therefore, we choose to fit the observational data for the two growth periods separately.

\vspace{2mm}
\section{Result}
\label{sec:result}

\begin{figure*}
\begin{center}
\includegraphics[width=83mm]{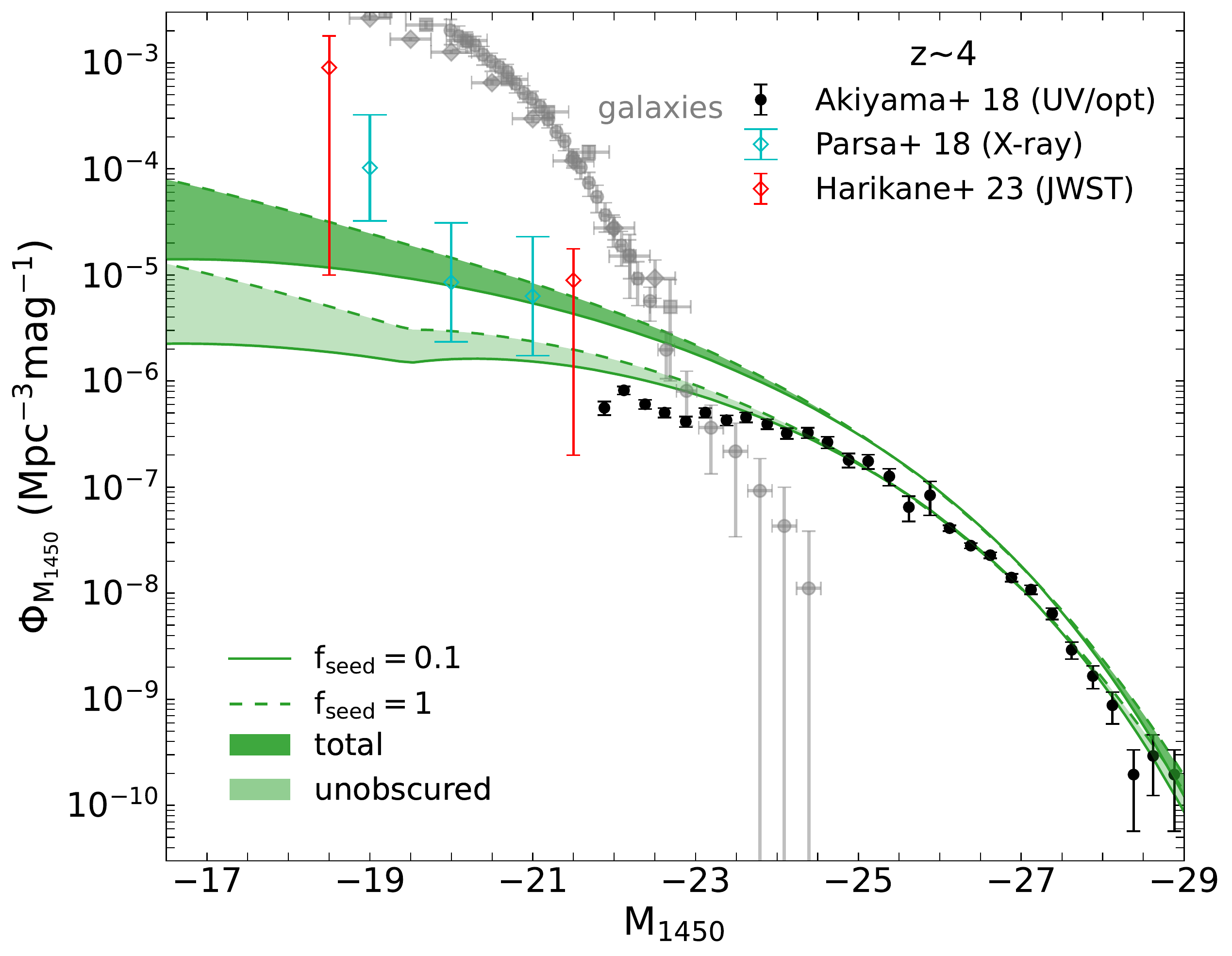}\hspace{5mm}
\includegraphics[width=83mm]{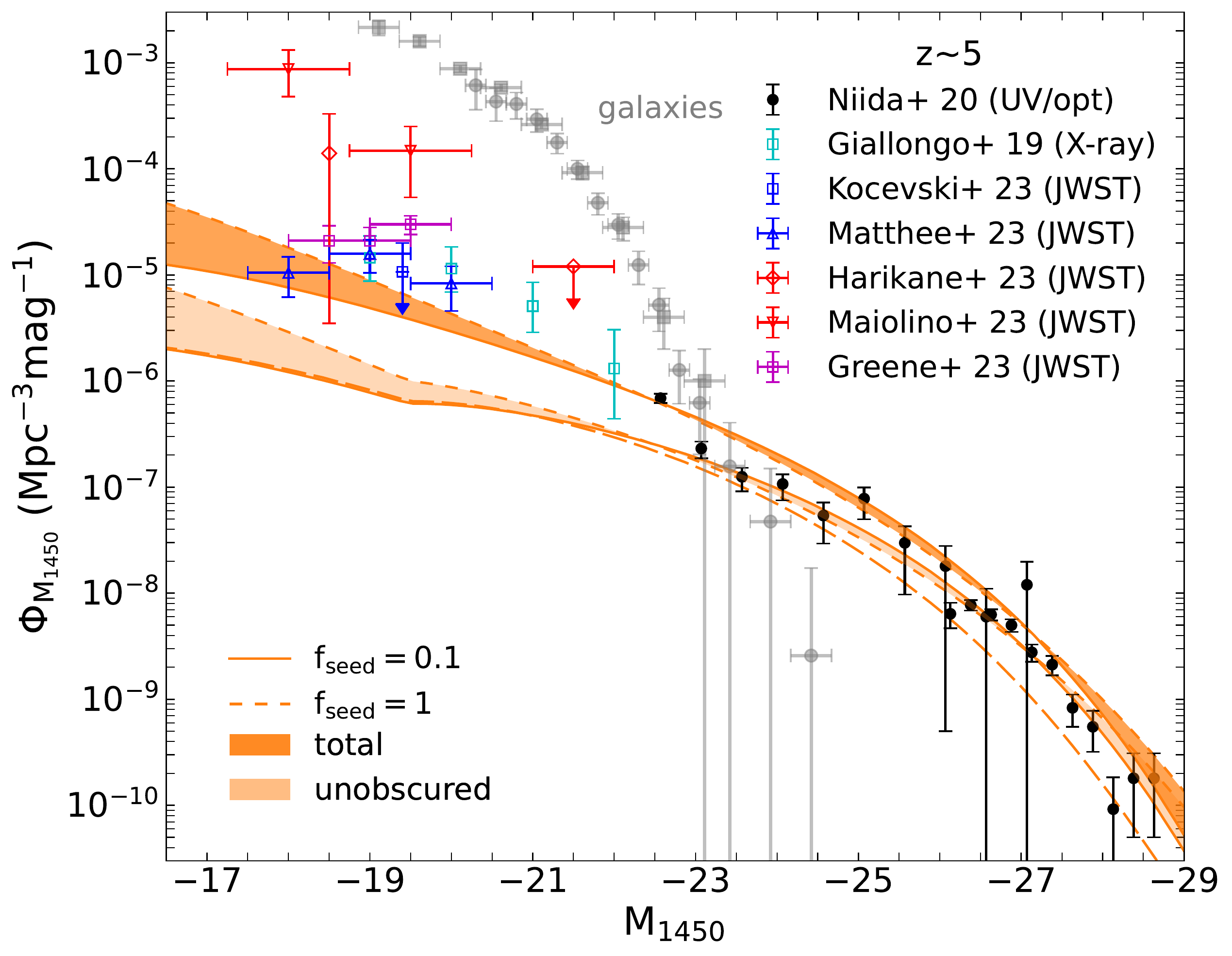}\\
\includegraphics[width=83mm]{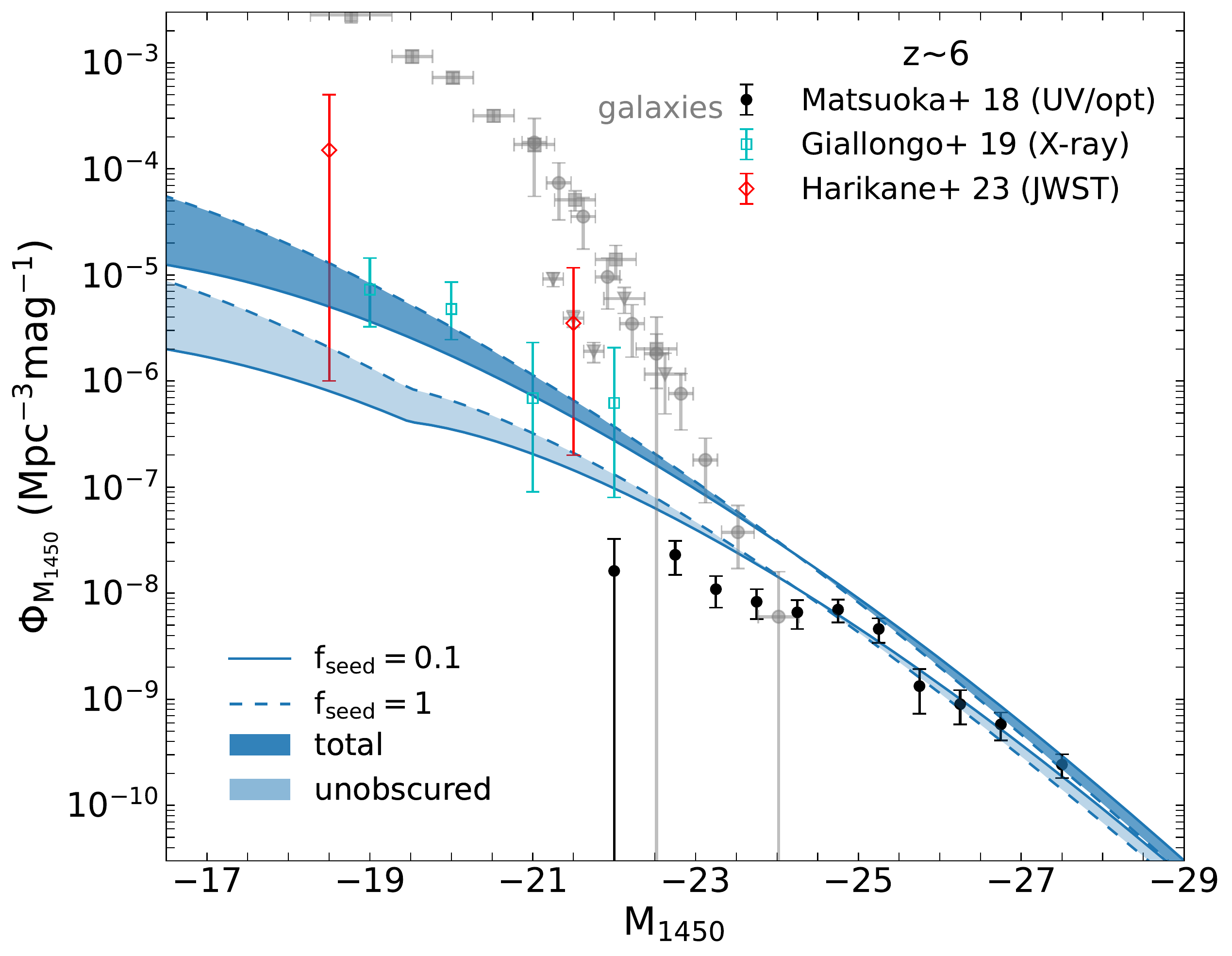}\hspace{5mm}
\includegraphics[width=83mm]{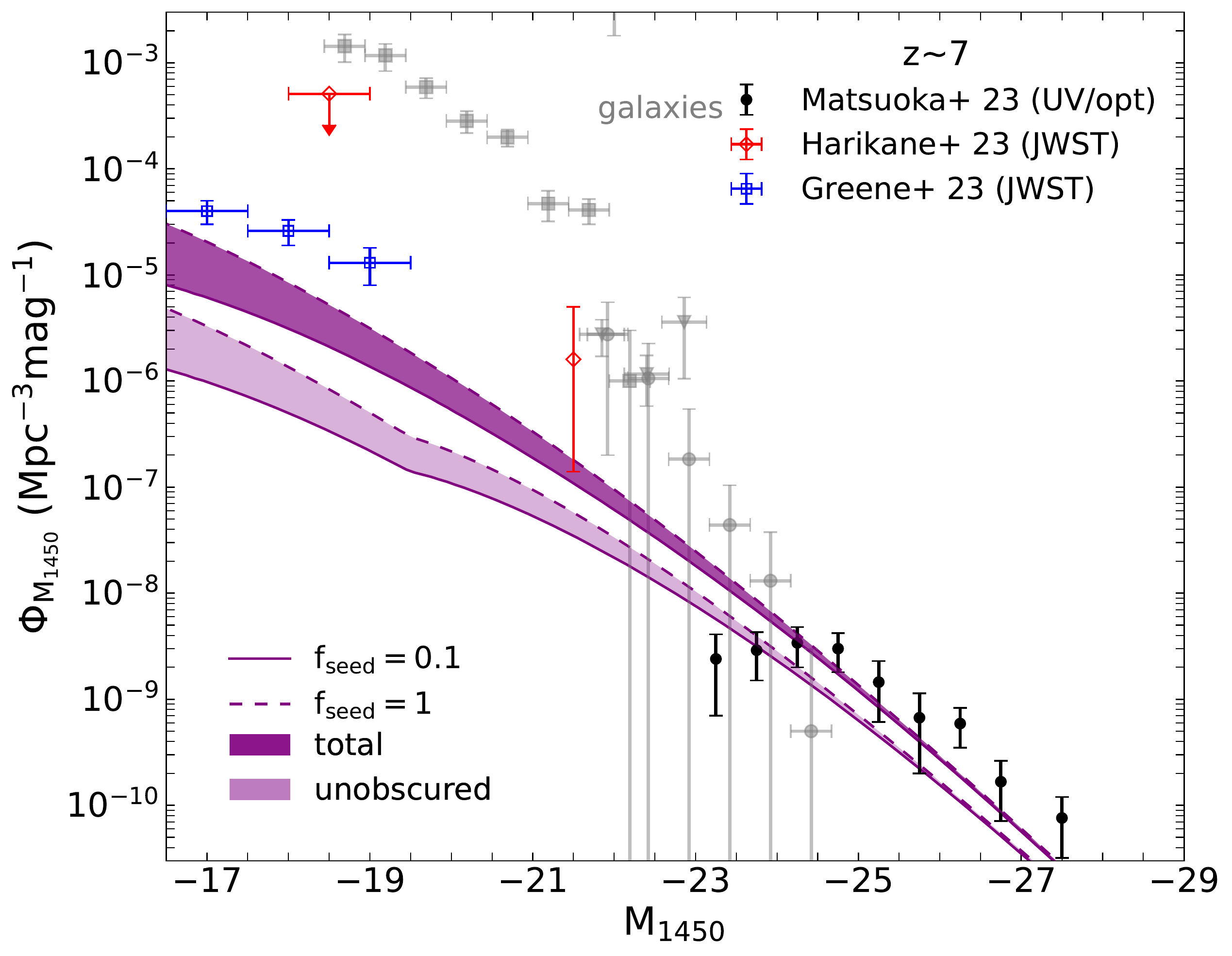}
\caption{Quasar luminosity functions at $4\lesssim z \lesssim 7$ derived from the best-fit model parameters.
The unobscured and total (obscured+unobscured) population are shown with light and dark shaded regions. 
The solid and dashed curves represent the models for the cases with $\fseed=0.1$ and $\fseed=1.0$.
The long dashed curve in the $z=5$ panel shows the QLF model in the case where only the $z\simeq 6$ and $z\simeq 4$ QLF data 
are used for the parameter fitting in the fiducial case.
The QLF data constructed by UV-optical surveys are adopted from \citet{Akiyama_2018}, \citet{Niida_2020},
\citet{Matsuoka_2018}, \citet{Matsuoka_2023}, and the data points based on X-ray observations taken from 
\citet{Parsa2018} and \citet{Giallongo2019}. 
The faint end of the QLF is also constrained by the discovery of low-luminosity AGNs via JWST 
\citep{Kocevski_2023,Harikane_2023,Matthee_2023,Maiolino_2023Jades,Greene_2024}, where the total UV magnitudes of the galaxy+AGN sources are adopted.
The galaxy luminosity function data are shown with grey symbols taken from 
\cite{Harikane_2022a}, \cite{Bouwens_2021}, \cite{Finkelstein_2015}, and \cite{Bowler_2015,Bowler_2017}, 
respectively.
}
\label{fig:LF_z}
\end{center}
\vspace{4mm}
\end{figure*}

\subsection{Parameter optimization}
In Table~\ref{tab:para}, we summarize the parameters optimized for reproducing the observed QLFs at $4\leq z\leq 6$, as well as the BHMF at $z=6$.
The results include the peak values of their one-dimensional posterior distribution,
as well as the 16\%, 50\%, and 84\% quantiles in the cumulative distribution in the parentheses.
As shown in the best-fit results, the intrinsic Schechter-like ERDF evolves across different redshift intervals.
Specifically, the power-law slope $\alpha$ increases and the characteristic Eddington ratio $\lambda_0$ decreases toward lower redshifts.
The tendency appears to be consistent with what is seen in the observed ERDF between $z=4$ and $z=6$ (see \citealt{He_2023}).
The duration of quasar activity also extends at lower redshifts, estimated as $\tau \simeq {\rm a~few}\times 100$ Myr within the $4<z<6$ range,
while $\tau \sim 20$ Myr at $z>6$.
This result suggests that accreting BHs change their growth speeds (i.e., their Eddington ratios) once or twice during this period at $4<z<6$.
Our phenomenological model also indicates a larger value of $\delta$ at $z<6$ that induces substantial suppression of 
mass growth for heavier SMBHs with $M_\bullet >10^8~\msun$.
The deceleration and deviation from the exponential growth manner lead to 
the downsizing or anti-hierarchical evolution of massive BH populations \citep{Ueda_2014,He_2023}.

The best-fit parameters in our model are connected to the fraction of seed BHs $\fseed$, a key variable representing the total number of BHs 
that contribute to the assembly of SMBHs in quasar host galaxies by $z\simeq 6$.
With the smaller value of $\fseed=0.1$ ($N_{\rm seed}\simeq 10^{-4}~{\rm Mpc}^{-3}$), our model requires a higher average speed of BH growth 
(primarily driven by larger values of $\lambda_0$ and $\alpha$) to align with the bright end of the QLF data.
In this scenario, fewer quasars are populated at the fainter end of the QLF, leading to a flatter slope at the faint end.
On the other hand, with the higher value of $\fseed=1.0$ ($N_{\rm seed}\simeq 10^{-3}~{\rm Mpc}^{-3}$), the model indicates that BHs typically 
grow at moderate speeds but only a small fraction of these BHs experience significant mass growth.
This case maintains a steeper slope in both the QLF and BHMF at the fainter and lower-mass ends.

\subsection{Evolution of quasar luminosity functions}
In Figure~\ref{fig:LF_z}, we present the theoretical model of QLFs at various redshifts of $z=4-7$.
The QLFs for the unobscured and total population are shown with the light and dark colored curves.
The two cases with different seeding fractions are represented with the solid ($\fseed=0.1$) and dashed ($\fseed=1.0$) curves. 
The QLF data are available on GitHub\footnote{\texttt{QLF} database: \url{https://github.com/WenxiuLiii/QLF}.}.
We also present the QLF model at $z=5$ calibrated only by the $z=6$ and $z=4$ observed QLF data with the long-dashed curve.
The result shows a relatively large offset from the binned QLF data.

Overall, our best-fit model for the unobscured QLFs agrees with the constraints from the rest-UV based QLF observations at $-29\lesssim \Muv \lesssim -24$
\citep{Akiyama_2018,Niida_2020,Matsuoka_2018,Matsuoka_2023}, 
while the model exhibits a moderate discrepancy from the observational data at the fainter end. 
The QLF model for the total population, including both unobscured and obscured quasars, is broadly consistent 
with the abundance of the X-ray selected faint quasars at $\Muv>-22$ \citep{Parsa2018,Giallongo2019}.
This abundance is notably higher than what is expected from extrapolating the rest-UV-based unobscured QLF to the faint end.

For comparison, we also incorporate the observed abundances of faint broad-line AGNs at $z\sim 4-7$ identified through JWST observations; 
unobscured AGNs \citep{Harikane_2023,Kocevski_2023,Maiolino_2023Jades} and dust-reddened obscured AGNs, so-called little red dots \citep{Matthee_2023,Greene_2024}.
Their results provided constraints on the abundance of these faint quasars, with values ranging in $\Phi_{\Muv}\simeq 10^{-5} -10^{-3}$ Mpc$^{-3}$ mag$^{-1}$ 
at $\Muv \sim -18$, where the total UV luminosity includes both the host galaxy and AGN contributions. 
Our QLF model for the total population (darker-colored shaded regions) explains well the abundance of 
these dust-reddened AGNs at $z\sim 5$ and $z\sim 7$, as it aligns with the abundance of X-ray selected obscured AGNs at $4<z<6$.
These agreements show the robustness of our model in capturing the key characteristics of the AGN population across a broad range of 
observational methodologies.

On the other hand, it is worth noting that dust-dereddened bolometric corrections of those red AGNs are dependent on their intrinsic
spectral energy distribution, which introduces significant uncertainties in calculating bolometric luminosities.
Assuming the low-redshift quasar composite spectrum \citep{VandenBerk_2001} and a dust extinction law \citep{Calzetti_2000},
the AGN bolometric luminosity functions have been constructed \citep{Greene_2024,Kokorev_2024}.
Their results reveal AGN abundances more than one order of magnitude higher compared to those observed in X-ray and UV studies
for $L_{\rm bol}>10^{46}~{\rm erg~s}^{-1}$, as detailed in Fig.~1 of \citet{Inayoshi_2024}.
Although such a discrepancy poses a challenge to most theoretical models, the determination of the luminosity function shape for
those reddened AGNs requires a more extensive sample of spectroscopically confirmed broad-line AGNs.
Additionally, photometric data at longer-wavelength (e.g., MIRI) are essential for a better estimation of the bolometric luminosity.
%

The diversity in JWST-identified AGN abundance measurements across different studies highlights significant discrepancies.
\cite{Maiolino_2023Jades} and \citet{Harikane_2023} reported a $5-10~\%$ presence of AGNs in star-forming galaxies at $z\simeq 4-7$,
implying an AGN abundance of $\Phi_{\Muv}\gtrsim 10^{-4}-10^{-3}~{\rm Mpc}^{-3}$ at the faint end.
This estimate is $1-2$ orders of magnitude higher than those derived from other JWST-based AGN studies using the observed survey volume density 
for calculations \citep[e.g.,][]{Kocevski_2023, Matthee_2023,Greene_2024}.
The differences in abundance estimates may stem from the complexities and uncertainties associated with the selection function of JWST NIRSpec targets, 
especially for slit spectroscopy modes influenced by the sample pre-selection criteria in each survey. 
This issue on the analysis has been noted in \cite{Harikane_2023}, where AGN abundances derived from the ratio of the number of detected AGNs to the survey volume are in closer agreement with the results based on the same method, yielding a lower bound of the abundance.

The choice of the seeding fraction impacts quasar abundance at the fainter end of $\Muv > -22$, 
though the theoretical QLFs remain almost consistent at the brighter end over $z=4-7$.
Even with a tenfold difference in the seeding fraction, the abundance for both unobscured and total populations 
at the faint end increases only by a factor of $\sim 2-4$ across all redshifts, when comparing scenarios with $\fseed=0.1$ and $\fseed=1.0$.
This smaller increase in number, as opposed to a direct tenfold enhancement, results from adjustment of BH growth model parameters
constrained by the QLF observations in the bright end.
Intriguingly, with the higher seeding fraction, the theoretical QLF model for the total BH population at $z\simeq 5$
(predominantly obscured AGNs) shows better agreement with the abundance of dust-reddened broad-line AGNs found in 
the JWST observation programs \citep{Matthee_2023}.
Further constraints on the faint AGN abundance will improve our understanding to the BH seed formation and growth more accurately
\citep{Inayoshi_ARAA_2020,Volonteri_2021} and the impact of these AGNs on cosmic reionization \citep[e.g.,][]{Fontanot_2020,Dayal_2024}.

\begin{figure*}
\centering
\includegraphics[width=83mm]{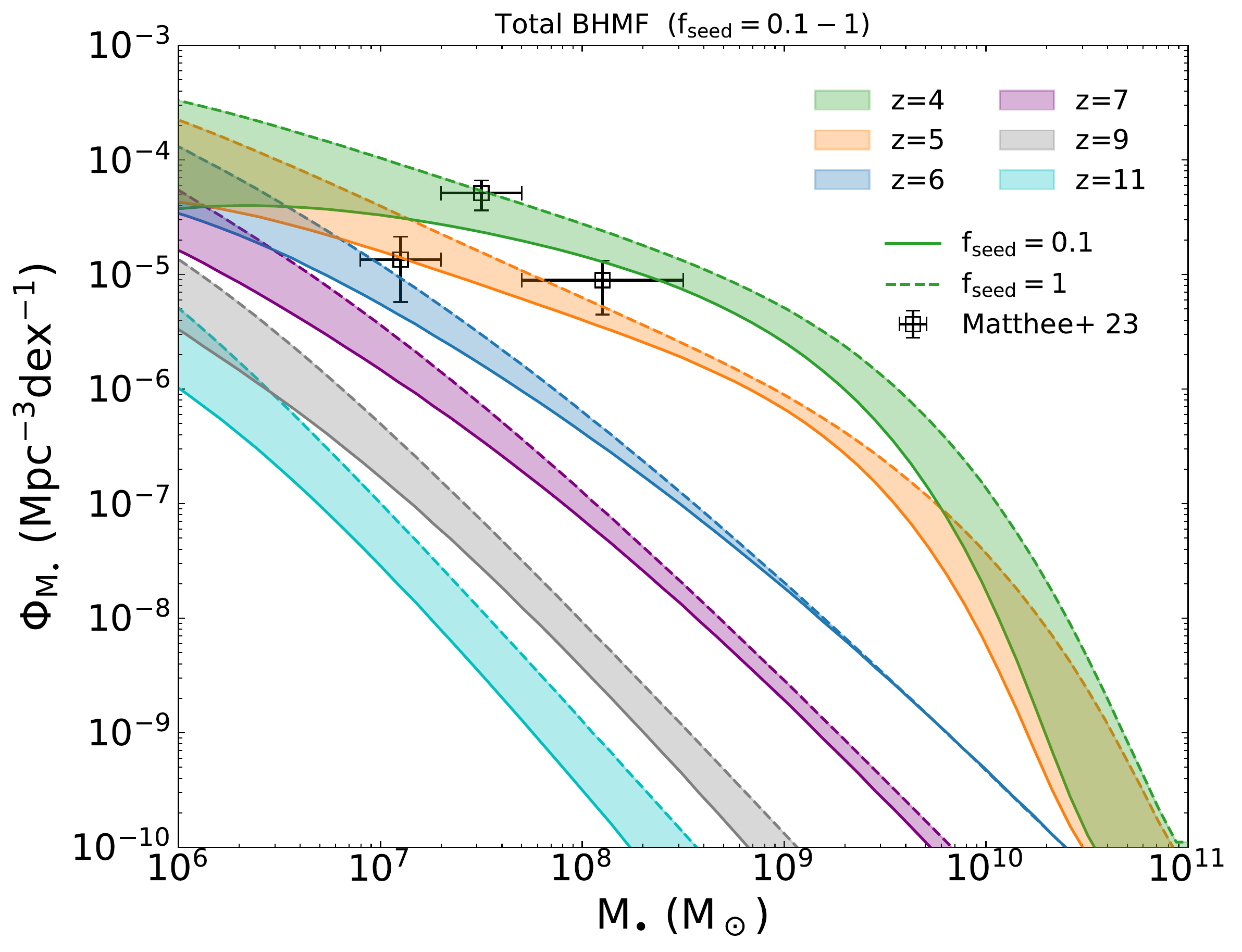}\hspace{5mm}
\includegraphics[width=83mm]{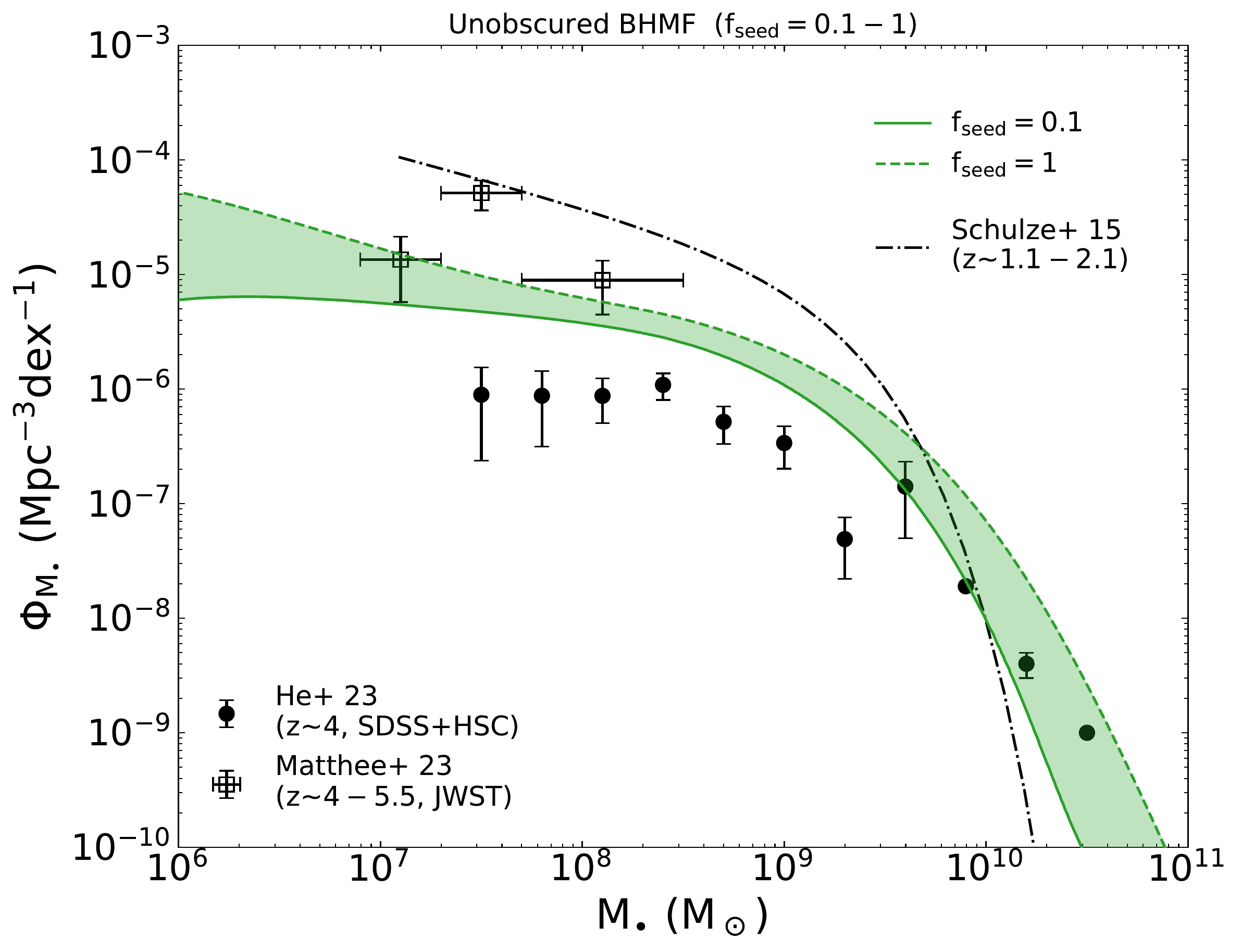}
\caption{{\it Left}: Predicted black hole mass functions for the total (unobscured + obscured) population at $4 \lesssim z \lesssim 11$ 
for the cases with $\fseed=0.1$ (solid) and 1.0 (dashed).
At $z\lesssim 5$, the growth of massive BHs with $\mbh >10^9~\msun$ is decelerated, leading to a steeper slope at the high-mass end. 
The theoretical QLF at $z=4$ is consistent with the binned BHMF constructed from dust-reddened, broad-line AGNs found by JWST observations 
\citep[][black squares]{Matthee_2023}.
{\it Right}: Black hole mass functions for the unobscured population at $z=4$, for the cases with $\fseed=0.1$ (solid) and 1.0 (dashed).
For comparison, we present the unobscured BHMF reconstructed by wide-field surveys at $z\sim4$ \citep[][black circles]{He_2023}
and at lower redshifts \citep[][black curve]{Schulze_2015}.
}
\label{fig:MF}
\vspace{4mm}
\end{figure*}

\subsection{Evolution of black hole mass functions}
In the left panel of Figure~\ref{fig:MF}, we present our theoretical model of BHMFs at $z=4-11$ for the total BH population 
including both unobscured and obscured AGNs.
The two cases with different seeding fractions are represented with the solid ($\fseed=0.1$) and dashed ($\fseed=1.0$) curves.
For both seeding fractions, the BHMF evolves preferentially at the high mass end with $\mbh>10^8~\msun$ between redshifts of $z=7$ and 5.
This rapid growth of high-mass BHs is linked with the QLF evolution at $\Muv \lesssim -23$,
corresponding to the Eddington luminosity for BHs with $\mbh\gtrsim 7\times 10^7~\msun$.
Toward a lower redshift at $z\sim 4$, the growth of these massive BHs slows down, resulting in a sharp cutoff in the BHMF 
at masses of $\mbh\gtrsim 10^{10}~\msun$.

We also overlay the BHMF at $z=4.2 -5.5$ in the low-mass regime of $10^7 \lesssim M_\bullet/\msun \lesssim 10^8$
constructed from the dust-reddened, broad-line AGNs \citep{Matthee_2023}\footnote{
The binned BHMF from these obscured AGNs does not account for sample incompleteness at the low-mass end.}.
Their findings are generally consistent with our predicted BHMF shape at $z=4$ across a seeding fraction range of $\fseed=0.1-1.0$,
despite the fact that we do not use these data for the parameter fitting.
This alignment validates our theoretical model in representing the broader BH population.
More complete samples in further observations enable us to constrain the BH seeding and growth model, primarily regarding the cosmic BH abundance 
at birth, $N_{\rm seed}\sim 10^{-3}(\fseed/1.0) ~{\rm Mpc}^{-3}$.

In the right panel of Figure~\ref{fig:MF}, we present the predicted BHMF for unobscured AGNs at $z=4$ with $\fseed=0.1$ (solid) and 1.0 (dashed),
along with the observed BHMFs for unobscured AGNs at $z\sim 4$ \citep[][black circles]{He_2023} and for dust-reddened AGNs \citep[][black squares]{Matthee_2023}.
For both the seeding fractions, our BHMF prediction is overall consistent with the data of \citet{He_2023} but shows 
a larger offset in the low-mass end.
The abundance of unobscured BHs at $\mbh >10^8~\msun$ reaches $\Phi_{\mbh}\simeq 4\times 10^{-6}~{\rm Mpc}^{-3}{\rm dex}^{-1}$ 
at $z\simeq 4$, nearly $\sim 10\%$ of the BH abundance in the same mass range at lower redshifts of $z<2$ \citep{Schulze_2015}.

As a caveat, our model takes into account BH populations formed in relatively biased regions of the universe with mass variance of $\gtrsim 3\sigma$,
neglecting contributions from substantially low-mass BH populations and their mass growth.
Therefore, the predicted BHMF provides a lower bound at low mass regimes of $\mbh<10^{7}$ -- $10^8~\msun$.
The propeties of these low-mass BHs will be probed by space-based gravitational-wave detectors such as the Laser Interferometer Space Antenna 
(LISA; \citealt{Amaro-Seoane_2023}), Tian-Qin \citep{Luo_2016, Mei_2021}, and Taiji \citep{Ruan_2018,Ruan_2020}.

\subsection{The cosmic evolution of BH mass density}
\begin{figure}
\centering
\includegraphics[width=85mm]{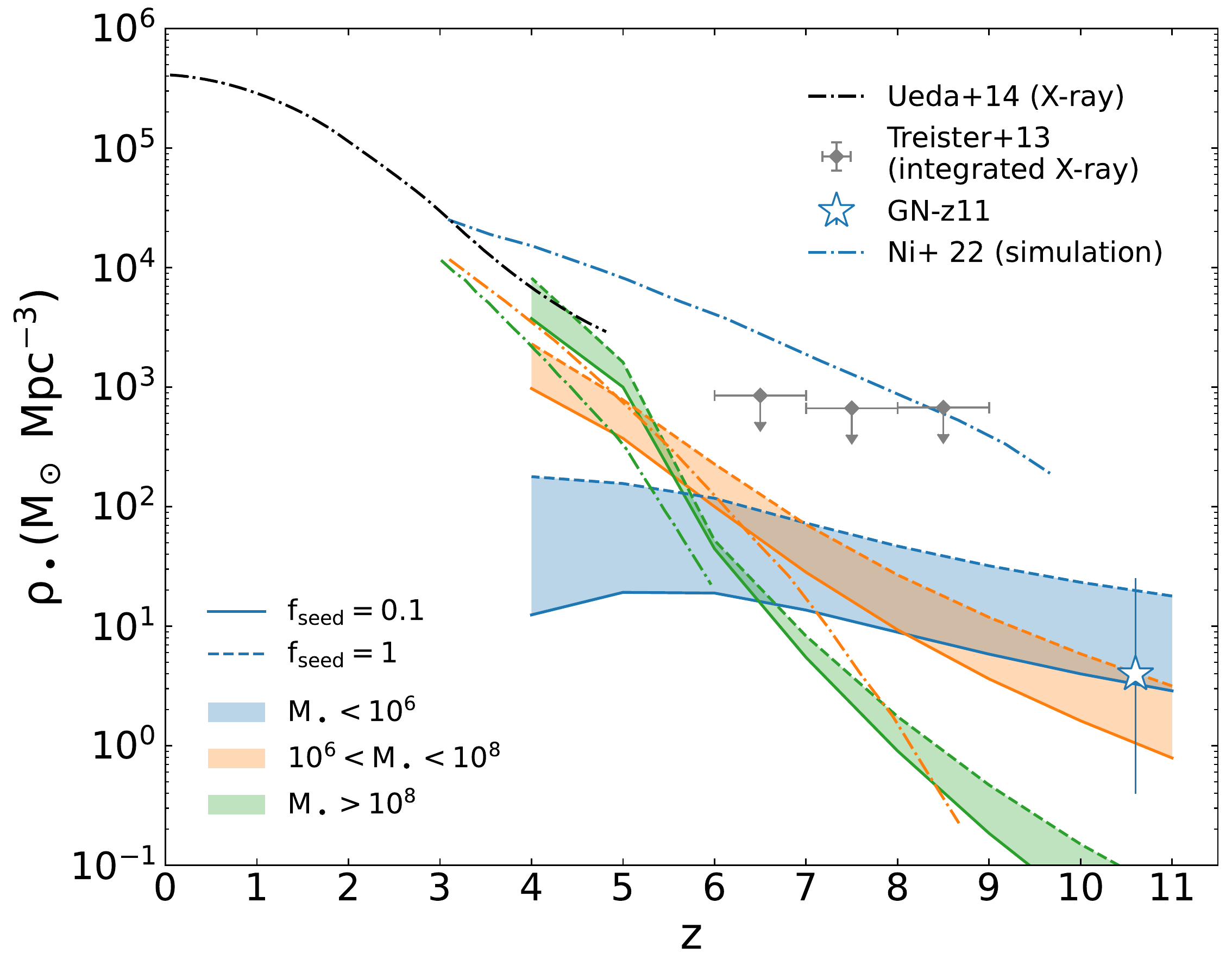}
\caption{Cosmic evolution of the BH mass density in a comoving volume at $4\leq z \leq 11$ at three mass ranges: 
$\mbh<10^6~\msun$ (blue),  $10^6<\mbh/\msun <10^8$ (orange), and $\mbh>10^8~\msun$ (green).
The lower and upper bound of the shaded areas are set to $\fseed=0.1$ (solid) and $\fseed=1.0$ (dashed).
At $z\lesssim 5$, 
the mass density of X-ray-selected AGNs with $L_{\rm bol} \gtrsim 10^{43}~{\rm erg~s}^{-1}$ is shown in the black dashed-dotted line \citep{Ueda_2014}.
The star symbol represents the value inferred from GN-z11,
a luminous galaxy at $z=10.6$, if an accreting BH with $\mbh \simeq 1.6 \times 10^6~\msun$ is hosted as reported 
by \citep{Maiolino_2023}.
The dashed-dotted curves present the results from the cosmological simulations by \cite{Ni2022}.
The upper bounds of the mass density accreted onto BHs measured by the cosmic X-ray background are shown with grey diamond symbols \citep{Treister_2013}.
}
\label{fig:rhoM_z}
\vspace{4mm}
\end{figure}

In Figure~\ref{fig:rhoM_z}, we present the evolution of the cumulative BH mass density $\rho_\bullet(z)$ 
within a comoving volume for the cases with $\fseed=0.1$ (solid) and 1.0 (dashed).
The cumulative BH mass density is derived by integrating the total BHMF shown in Figure~\ref{fig:MF}
\begin{equation}
 \rho_\bullet(z)=\int_{\mathcal{I}} \Phi_{\mbh} (z) \mbh ~\D \log \mbh.
\end{equation}
We consider three mass ranges: $\mbh \leq 10^6~\msun$ (blue)\footnote{The population includes BHs that participate 
in the assembly of SMBHs, but do not consider dormant seed BHs, which occupy $90\%$ of the total seeds 
formed for the case of $\fseed=0.1$ in our calculation.}, 
$10^6~\msun < \mbh < 10^8~\msun$ (orange), 
and $\mbh \geq 10^8~\msun$ (green).
At extremely high redshifts ($z\gtrsim 9$), the total BH mass density is constituted by the low-mass population
with $\mbh \lesssim 10^{6}~\msun$. 
As the universe approaches the end of cosmic reionization ($z\lesssim 6-7$), BHs with $\mbh > 10^6~\msun$
begin to account for the majority of the total BH mass density, 
resulting in the anti-hierarchical evolution of massive BHs.
At $z\lesssim 5$, the growth pace of the heaviest BHs with masses exceeding $10^8~\msun$ begins to decelerate.
By $z\sim 4$, the total cumulative density reaches $\rho_\bullet \simeq 5\times 10^3~\msun~\mpc^{-3}$ for $\fseed=0.1$,
and $\rho_\bullet \simeq 10^4~\msun~\mpc^{-3}$ for $\fseed=1.0$, respectively.
This value corresponds to more than $1\%$ of the total BH mass density observed in the present-day universe, 
$\rho_{\bullet, 0} \simeq (3-5) \times 10^5~\msun~\mpc^{-3}$ 
\citep[e.g.,][]{Marconi_2004,Shankar_2009,Vika_2009}.
The overall trends for the massive BH populations with $\mbh>10^6~\msun$ at $z<7$ are in good agreement with those 
obtained from the cosmological simulations in \citet{Ni2022} (dashed-dotted curves).


The unresolved fraction of cosmic X-ray background at $z \gtrsim 6$ places constraints on the global BH accretion history 
\citep{Salvaterra_2012, Treister_2013}.
Upper limits on the density of accreted mass (grey arrows) prevent an overproduction of massive BHs and their seeds at the cosmic dawn. 
At $z\sim 4-5$, our model can be cross-referenced with the accumulated BH mass density inferred from the abundance of X-ray selected AGNs
with $L_{\rm bol} \gtrsim 10^{43}~{\rm erg~s}^{-1}$ \citep[black dashed-dotted;][]{Ueda_2014}\footnote{
\citet{Ueda_2014} calculated the density of mass accreted onto BHs assuming a radiative efficiency of $\eta_0=0.05$. 
This is half of our fiducial value of $\eta_0=0.1$, increasing the mass density by a factor of two. 
Therefore, we have rescaled their results by a factor of 0.5 for comparison to our study.}.
Overall, our result in the range of $\fseed=0.1-1$ shows good agreement with those derived from X-ray observations.
We note that we have taken into account the contributions of BHs from relatively biased regions of the universe
with mass variance of $\gtrsim 3\sigma$, but overlooked migration of less massive seed BH populations originating in 
more typical regions, where galaxies form in parent halos with $\mh\lesssim 10^{11}~\msun$ at $z\sim 6$.

In the early evolution of low-mass BHs, our model presents a mass density at $z\sim11$ that is 
consistent with the observations of GN-z11, as inferred from its cosmic volume density and BH mass,
$\simeq 4.0^{+21.3}_{-3.6}~\msun~\mpc^{-3}$ (star symbol), considering the uncertainty in its mass and abundance measurement \citep{Maiolino_2023,Oesch_2016}. 
Intriguingly, even sophisticated cosmological simulations fail to attain this value, which likely points to 
their insufficient treatments of BH seeding \citep[dashed-dotted curves;][]{Ni2022}. 
This highlights the importance of our approach that connects BH seeding with subsequent growth and the unique insights into the understanding of BH evolution,
particularly in these early epochs of the universe.

\vspace{4mm}
\section{Discussion}
\label{sec:discussion}

\subsection{Individual BH growth}
\label{sec:discussion_a}

We explore the individual evolutionary tracks of BHs starting from their seeding epochs with
intermittent accretion rates. 
We generate a sample of $10^6$ BHs following their formation times and mass distribution at birth 
as described in \cite{Li_2021,Li_2023}.
Adopting the best-fit parameters for each redshift range (see Table.~\ref{tab:para} and Figure~\ref{fig:LF_z}) with the case of $\fseed=0.1$, we assign a single value of 
the Eddington ratio generated from the Schechter-like ERDF with $\lambda_0(z)$ and $\alpha(z)$ in each time interval of $\tau(z)$.
With these parameters, we grow the individual BHs until $z=4$ and study their statistical properties.
%

\begin{figure*}
\centering
\includegraphics[width=160mm]{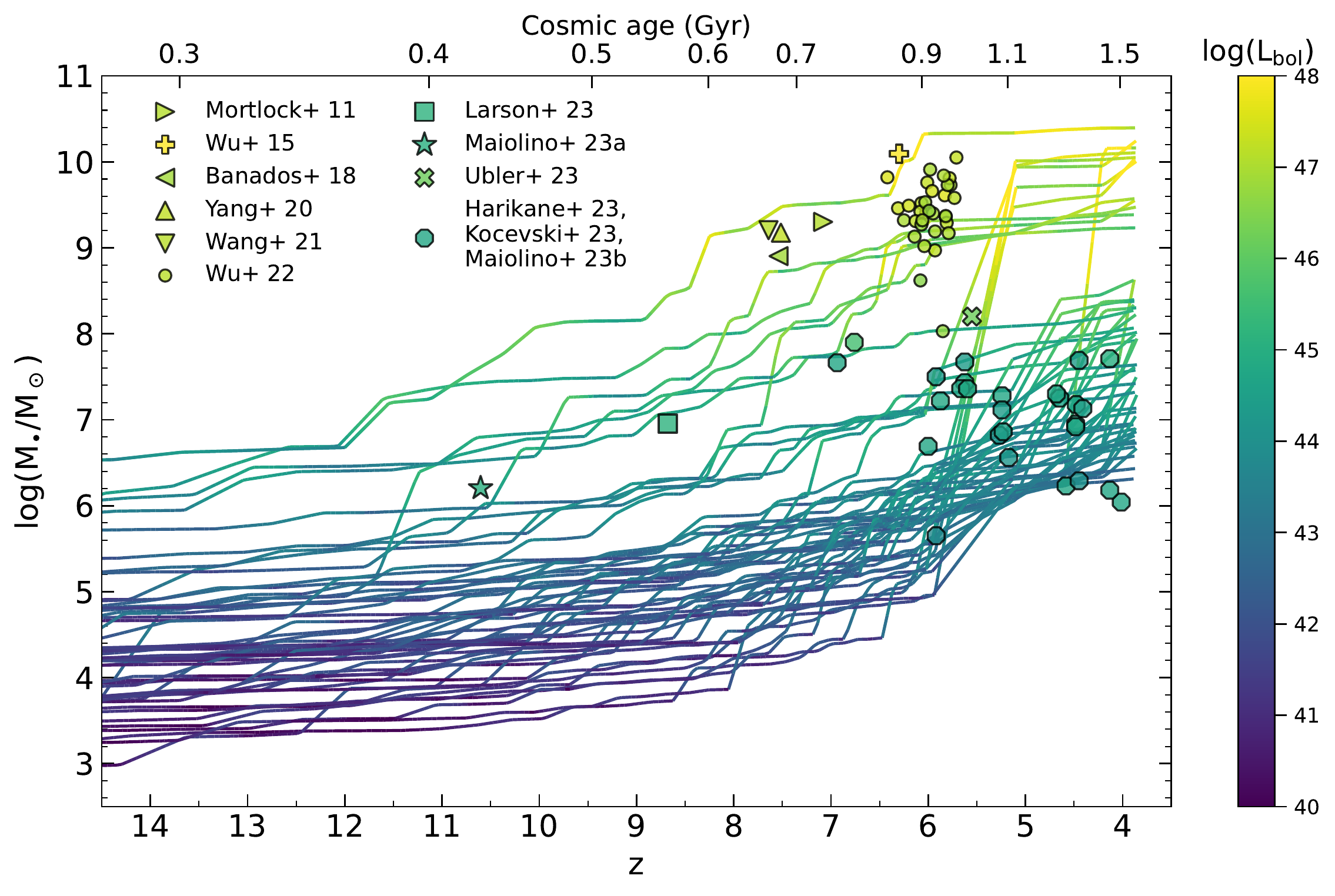}
\caption{
Growth tracks of individual BHs with the best-fit model parameters calibrated with the observed QLFs.
Among all the samples ($N=10^6$), we present three BH populations with colored lines: 
(i) those reaching $\mbh \geq 10^9~\msun$ by $z\gtrsim 6$, 
(ii) those with $10^6\leq \mbh/\msun \leq 10^8$ at $z=5$, and 
(iii) those have grown to $\mbh \geq 10^{10}~\msun$ at $z= 4$.
The color of each curve represents the bolometric luminosity for the accreting BH associated with its growth.
We overlay the observational data points of high-$z$ bright quasars previously observed at $z>6$ \citep{Mortlock_2011,Wu_2015,Banados_2018,Onoue_2019,Yang_2020,Wang_2021,Wu_2022} and those detected by JWST \citep{Onoue_2023,Kocevski_2023,Ubler_2023,Larson_2023,Harikane_2023,Maiolino_2023,Maiolino_2023Jades}.
}
\label{fig:Mevol}
\vspace{4mm}
\end{figure*}

Figure~\ref{fig:Mevol} presents the mass growth tracks between $z=15$ and $4$ for a randomly-selected subset of 
100 BHs out of the $10^6$ samples (grey curves).
We here employ color-coding to represent three distinct BH populations: 
(i) those reaching $\mbh \geq 10^9~\msun$ by $z\gtrsim 6$,
(ii) those with $10^6\leq \mbh/\msun \leq 10^8$ at $z=5$,
and (iii) those have grown to $\mbh \geq 10^{10}~\msun$ at $z= 4$.
For classes (i) and (iii), we highlight all tracks that meet the conditions among the whole $10^6$ samples, 
while for class (ii), we only present 40 curves out of an approximate total of 40,000 for illustrative purposes.
The color of each curve represents the bolometric luminosity for the accreting BH.
Additionally, we overlay the data points of high-$z$ quasars previously known and recently reported by JWST observations 
\citep{Mortlock_2011, Wu_2015,Banados_2018,Onoue_2019,Yang_2020,Wang_2021,Wu_2022,Kocevski_2023,Ubler_2023,Harikane_2023,Larson_2023, Maiolino_2023,Maiolino_2023Jades}.

In our sample of $10^6$ objects, six BHs are identified as bright quasars hosting massive BHs with 
$\mbh\geq 10^9~\msun$ at $z>6$ (class i).
During an actively growing phase, the bolometric luminosity rises and reaches levels comparable to the brightest 
quasars observed by SDSS and HSC (filled circles representing data taken from \citealt{Wu_2022} and \citealt{Onoue_2019}).
The growth trajectory of the most massive BH explains the existence of J0313-1806 at $z=7.642$ \citep{Wang_2021},
J1120+0641 at $z=7.10$ \citep{Mortlock_2011}, and J0100+2802 at $z=6.3$ \citep{Wu_2015}, 
while that of the second-most massive BHs accounts for J1342+0928 at $z=7.54$ \citep{Banados_2018}.
Furthermore, the very distant quasar reported at $z=8.68$ by JWST/CEERS programs \citep[CEERS-1019;][]{Larson_2023}
and GN-z11 at $z=10.6$ by JWST/JADES programs \citep{Maiolino_2023} can be 
attributed to the BH population that ends up in typical bright quasars hosting $\mbh \gtrsim 10^8~\msun$ at $z\sim 6$.
Recent theoretical studies provide a formation model of those BHs in CEERS-1019 and GN-z11 based on heavy seed BHs
with $\mbh\gtrsim 10^{3}$ -- $10^{5}~\msun$ as shown in Figure~9 of \citealt{Li_2023} (see also an alternative model presented
in \citealt{Schneider_2023}).

A larger population of the BHs follow moderate growth tracks, where their masses lie in the range of $10^6\leq \mbh/\msun \leq 10^8$ at $z=5$ (class ii).
These low-luminosity AGNs with $L_{\rm bol}\lesssim 10^{45}~{\rm erg~s}^{-1}$ are more representative of the normal BH population 
rather than the ultra-rare and luminous quasars discussed above.
The successful spectroscopic identification of low-luminosity broad-line AGNs at $z=4$ -- 7 opens up a new parameter space for high-redshift AGN studies, thanks to the unprecedented infrared sensitivity of JWST.
Here, we overlay the spectroscopically-confirmed AGNs with JWST \citep{Kocevski_2023, Ubler_2023, Harikane_2023, Maiolino_2023Jades}.
This population can be explained by seed BHs with $\mbh \sim 10^{3}$ -- $10^{5}~\msun$ at $z\gtrsim 15$, originating from massive stellar remnants
formed in moderately-biased regions of the high-$z$ universe \citep[e.g.,][]{Valiante_2018,Lupi_2021, Li_2021, Sassano_2021, Toyouchi_2023}.

\begin{figure}
\centering
\includegraphics[width=85mm]{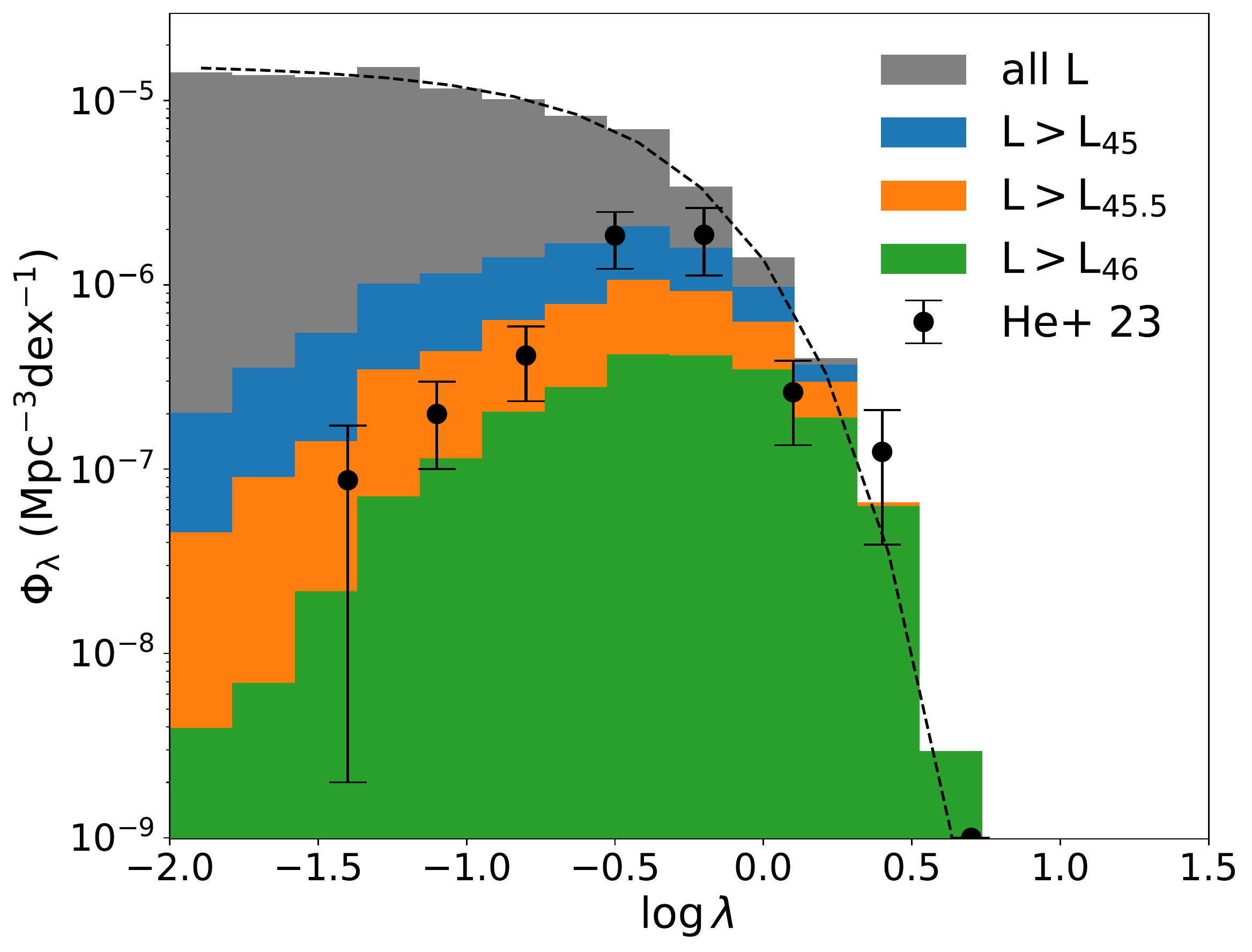}
\caption{Eddington ratio distribution functions of the AGN population at $z=4$ selected by imposing different detection limits of the quasar bolometric luminosity:
all BHs (grey), $L_\mathrm{bol} \geq L_{45}$ (blue), $L_{45.5}$ (orange), and $L_{46}$ (green).
The black symbols show the results from \citet{He_2023}, constructed from their $z\sim 4$ quasar sample with $L_\mathrm{bol} \gtrsim L_{45.5}$.}
\label{fig:lambdaSch}
\vspace{4mm}
\end{figure}

\begin{figure*}
\centering
\includegraphics[width=130mm]{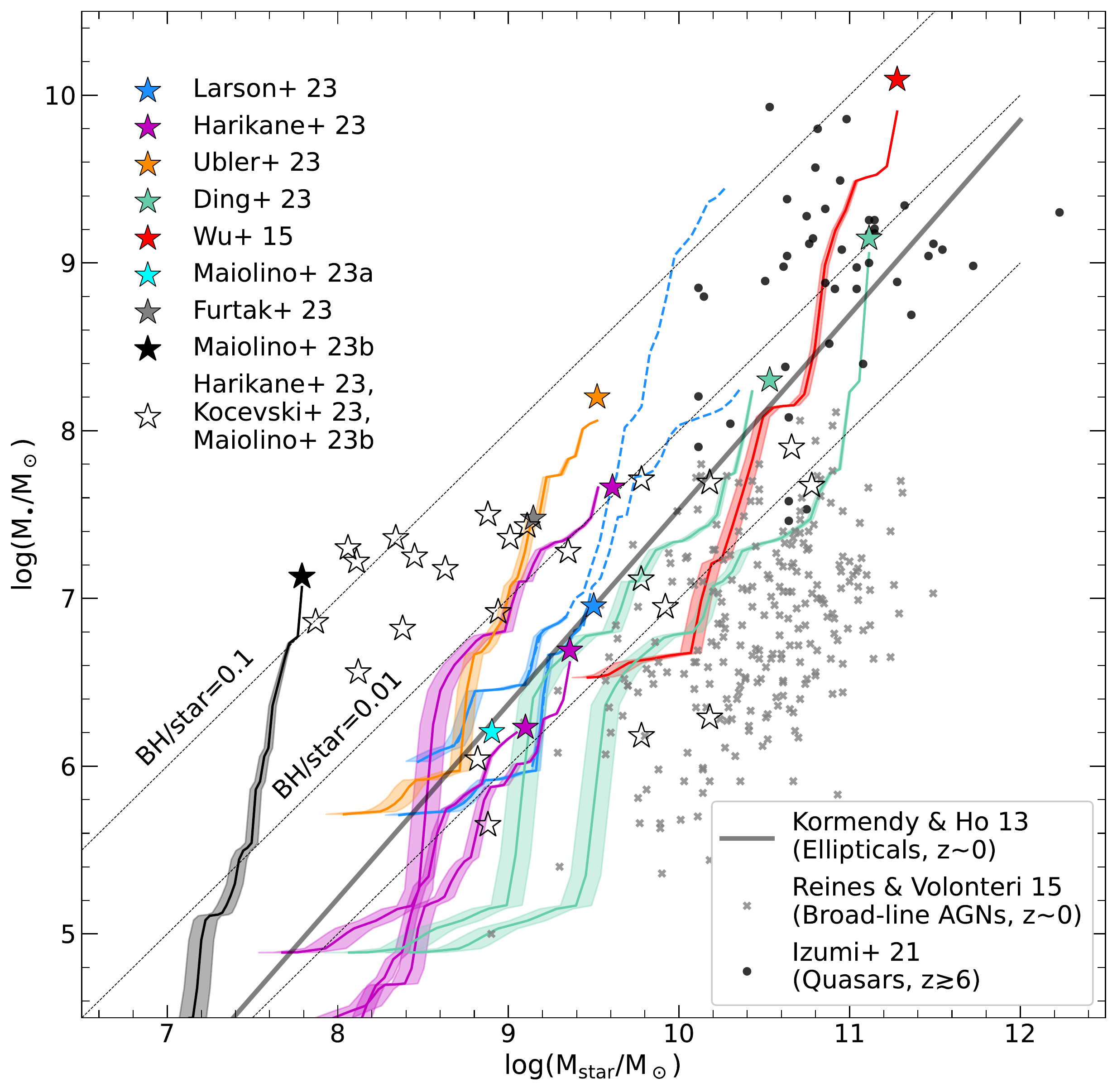}
\caption{
The predicted evolution of the BH-stellar mass relation, $\mbh$ -- $\mstar$, for quasars at high redshifts of $4<z<11$,
for which the virial BH mass and stellar mass (or gas dynamical mass with the [C~{\sc ii}] 158 $\micron$ line)
are measured; the $z\sim 8.68$ AGN \citep{Larson_2023}, two quasars with notably overmassive BHs \citep{Ubler_2023, Wu_2015}, 
two quasars J2236+0032 and J2255+0251 \citep{Ding_2023},
three quasars (GLASS 150029, CEERS 00397, and CEERS 00717; \citealt{Harikane_2023}), 
one overmassive quasar found by \citet{Maiolino_2023Jades}, 
and the BH in GN-z11 \citep{Maiolino_2023}.
Open star symbols represent the other low-luminosity AGNs at $z\sim 4$ -- 7 reported by JWST observations 
\citep{Kocevski_2023,Harikane_2023,Maiolino_2023Jades}.
The grey and black dots present the distribution of $z\sim 0$ broad-line AGNs \citep{RV_2015} and $z\sim 6$ quasars \citep{Izumi_2021},
where the total stellar mass and gas dynamical mass are applied, respectively.
The solid thick line shows the empirical BH-bulge mass relation seen in the nearby universe \citep{Kormendy_Ho_2013}, while
the diagonal dashed lines denote constant values of $\mbh/\mstar=0.1$, $0.01$, and $10^{-3}$.
}
\label{fig:Msigma}
\vspace{4mm}
\end{figure*}

Moreover, the BH growth model successfully reproduces the presence of 
the heaviest BHs, which have $\mbh \simeq 10^{10}\msun$ at $z\sim 4$ (class iii). 
These ultra-massive BHs originate from two distinct sources. 
The first origin is the brightest quasars, where the BHs have already achieved significant mass by $z\sim 6$ (class i). 
The second origin comprises rapidly accreting BHs, which have $\mbh \sim 10^5$ -- $10^7~\msun$ at $z\sim 6$. 
In the latter scenario, a small fraction of those normal BH populations at $z\sim 6$ undergo substantial mass accretion.
This high growth rate persists for longer periods at $z<6$ owing to the redshift dependence of the typical active duration $\tau(z)$,
which increases at lower redshifts.
At $5\leq z <6$, the active duration is $\tau\simeq 195$ Myr, which is comparable to the cosmic time elapsed between the redshift interval ($\Delta t_{\rm H}\simeq 237$ Myr).
Thus, a minor population of BHs experience significant mass accretion through super-Eddington phases at rates
with $\lambda \sim 4-5$.
This type of transient super-Eddington accretion is expected to take place in high-$z$ massive galaxies \citep[e.g.,][]{Inayoshi_2022a},
and the nature of radiatively inefficient accretion due to photon trapping \citep{Abramowicz_1988, Watarai_2000} might 
account for high-$z$ quasars with small proximity zones that indicate short lifetimes $t_{\rm Q} \sim 10^6$ yr \citep{Eilers_2021}.

As we shift toward lower redshifts, the growth of such a rare BH population is suppressed and thus their mass 
saturates at $\mbh \sim 10^9$ -- $10^{10}~\msun$.
Intriguingly, this outcome aligns with observations showing that the largest SMBHs have a mass of $\mbh \simeq 10^{10}~\msun$, 
nearly independent of redshift from the local ($z\simeq 0$) to the early ($z\gtrsim 6$) universe \citep{Inayoshi_Haiman_2016, King_2016}. 
Various physical processes would contribute to this outcome, including accretion disk instability \citep[e.g.,][]{Pringle1991,Yu2005}, 
and strong feedback mechanisms through outflows and jets counteracting the inflow motion \citep[e.g.,][]{Blandford_Begelman_1999,Ho2002,Ichikawa2017}.

\subsection{Observed ERDF}
\label{sec:discussion_b}
The observed quasar sample is biased toward bolometric luminosities above the detection limits of quasar surveys.
We study the observed ERDFs by imposing different detection limits to the bolometric luminosities of the quasar sample we generate in Section~\ref{sec:discussion_a}
with the BH growth parameters for the case with $\fseed=0.1$.

In Figure~\ref{fig:lambdaSch}, we present the intrinsic Schechter shape of ERDF for the whole BH sample following Eq.~(\ref{eq:Pl}),
along with the observed ERDF for quasars selected with bolometric luminosities $L_\mathrm{bol} \geq L_{45}=10^{45}$, $L_{45.5}=10^{45.5}$, and $L_{46}=10^{46}~\ergs$ at $z=4$.
The luminosities mimick the detection depths of quasar surveys.
The ERDF shape with a detection threshold is skewed to a log-normal one \citep{Willott_2010,Kelly_Shen_2013,Schulze_2015} and is consistent with the observed result of \citet{He_2023},
where their $z\sim 4$ quasar samples are collected with a detection limit of $L_\mathrm{bol} \gtrsim L_{45.5}$.
Our results indicate a large fraction of the BHs accreting with low-$\lambda$ values remain yet to be detected.
The agreement between the best-fit model and observation additionally supports that our model assumption is reasonable in reproducing the properties of the quasar population.

\subsection{Connection between BH and galaxy growth}
The assembly of the mass correlation between high-$z$ BHs and their host galaxies 
is crucial for understanding the establishment of their local relation \citep[e.g.,][]{Kormendy_Ho_2013},
despite its origin remaining one of the longest-standing unsolved puzzles in astrophysics
\citep[e.g.,][]{Murray2005,Silk2013,Cen2015,Ni2022,Inayoshi_2022a,Habouzit_2022}.
Direct measurements of the stellar mass of high-$z$ quasar hosts can be challenging without observations of rest-frame optical light.
In certain cases, the gas dynamical mass derived from [C~{\sc ii}] 158 $\micron$ serves as a proxy of the stellar mass. 
The measured BH-to-stellar mass ratio for luminous quasars significantly exceeds the local relation 
\citep[e.g.,][]{Wang2013,Venemans2017,Neeleman2021}.
In contrast, for low-luminosity quasars, the ratio aligns more close or even falls below the local relation \citep{Izumi_2019,Izumi_2021}.
In Figure~\ref{fig:Msigma}, the BH mass is presented against the gas dynamical mass for those AGN samples at $z\gtrsim 6$ 
(black circle, the data compiled by \citealt{Izumi_2021}).

Recent JWST observations have provided new insights into the evolution of the BH-to-stellar mass correlation 
through the discovery of low-luminosity AGNs in the high-$z$ universe and detection of stellar optical light in the rest frame 
\citep[e.g.,][]{Onoue_2023,Ding_2023}. 
Spectroscopic follow-up observations have confirmed 12 broad-line AGNs, allowing for the measurement 
of their BH masses using the H$\alpha$(H$\beta$)-based single-epoch method: 
CEERS 1670 at $z=5.24$ \citep{Kocevski_2023}, GS 3073 at $z=5.55$ \citep{Ubler_2023}, CEERS 1019 at $z=8.68$ \citep{Larson_2023}, 
and several additional sources at $z\sim 4-7$ \citep{Harikane_2023}. 
In addition, GN-z11, an extraordinarily luminous galaxy at $z=10.6$, has been reported to host an AGN, 
as evidenced by the detection of the high ionization [Ne~{\sc iv}] $\lambda$2423 transition and semi-forbidden 
nebular lines tracing the clouds of the broad-line regions \citep{Maiolino_2023}.
\cite{Furtak_2023} conduct deep spectroscopic JWST/NIRSpec observations and confirm a red quasar at $z=7.05$,
measuring the BH mass to be $3\times 10^7~\msun$ by its triply images provided by the strong lensing of the galaxy cluster Abell 2744.
A recent study by \citet{Maiolino_2023Jades} report the discovery of twelve broad-line AGNs at $z\sim 4-7$ in the JWST JADES survey.
The stellar mass of these AGN host galaxies can be estimated through AGN-host image decomposition, spectral fitting, 
or a combination of the two \citep{Kocevski_2023, Ubler_2023,Larson_2023, Harikane_2023, Furtak_2023,Maiolino_2023,Maiolino_2023Jades}. 
As illustrated in Figure~\ref{fig:Msigma}, the resulting BH-to-stellar mass ratio (star symbols) is systematically higher 
than that found in AGN-host galaxies at $z\sim 0$ \citep[grey circles;][]{RV_2015}, but it is more consistent with
the relation observed in local elliptical galaxies \citep[solid black line;][]{Kormendy_Ho_2013}.

In the following discussion, we examine the evolutionary track of the BH-to-stellar mass ratio from seeding time 
to the observed epoch by selecting several AGNs with measured BH and stellar masses. 
We utilize individual growth tracks of our BH samples shown in Figure~\ref{fig:Mevol} to analyze the BH mass assembly, 
identifying the masses of the BH and its parent halo at the seeding redshift $z_i$.
Although the stellar mass assembly history of these BH hosts remains uncertain with current observations, 
the stellar mass can be approximated to follow that of the parent dark matter halo as
$\mstar(z) = \fstar \fb \mh(z)$, where $\fb (=\Omega_{\rm b}/\Omega_{\rm m}) =0.16$ is the baryon fraction 
and $\fstar$ is the star formation efficiency from gas into stars.
We here adopt a constant value of $\fstar$ without dependence on the redshift and halo properties.
This approximation is broadly consistent with cosmological simulations that trace the assembly of quasar host galaxies at high redshifts
\citep[e.g.,][]{Valentini_2021,Zhu_2022}\footnote{
The star formation efficiency is expected to be as low as $\fstar \lesssim 0.01$ in dark-matter halos with $\mh \lesssim 10^8~\msun$
\citep{McCaffrey_2023},
whose virial temperature is below the atomic cooling threshold of $\sim 10^4~\K$ at redshifts of interest.
In contrast, the efficiency increases to $\fstar> 0.1$ in more massive halos, as expected for bright $z\gtrsim 10$ galaxies detected with JWST
\citep[e.g.,][]{Inayoshi_2022c,Boylan-Kolchin_2023,Ferrara_2023,Mason_2023,Shen_2023}.}.
However, in comparison with empirical relations suggested by \cite{Behroozi_2019} (see their Figure 9), 
our treatment would give an overestimate of the stellar mass at higher redshifts and lower halo masses.
To take account of this uncertainty, we therefore examine three cases with $\fstar = 0.05$, $0.1$, and $0.2$, and 
demonstrate that the choice of $\fstar$ does not affect our discussion for the BH-to-stellar mass ratio.
The parent halo mass growth of a host galaxy follows a functional form of $\mh(z) \propto e^{-Bz}$ 
\citep{Wechsler_2002,Neistein2008,Fakhouri_2010}, 
resulting in a halo-mass growth rate of $d \ln \mh/dt \propto (1+z)^{5/2}$. 
This rate is based on the extended Press-Schechter formalism and consistent with fitting the individual halo growth 
in cosmological $N$-body simulations \citep{Dekel_2013}.

We select nine observed AGNs with measured BH and stellar masses, 
including the most distant AGN \citep{Larson_2023}, three quasars with notably overmassive BHs \citep{Ubler_2023, Wu_2015,Maiolino_2023Jades}, 
two quasars separated from the light of their host galaxies by decomposition of NIRCam image \citep{Ding_2023},
and three quasars chosen from the ten reported in \cite{Harikane_2023}: GLASS 150029, CEERS 00397, and CEERS 00717, 
all with moderate BH-to-stellar mass ratios.
All of these quasar hosts have stellar mass measurement, except that the quasar discovered by \cite{Wu_2015} outshines its host galaxy,
for which we adopt the dynamical mass measured by the [C~{\sc ii}] 158 $\micron$ line \citep{Wang_2019}.
Here we note that a population of high-$z$ galaxies are recently unveiled with a stellar mass lower than the dynamical mass for up to 1 dex,
possibly indicating a large fraction of mass in gas phase \citep{deGraaff_2023}.
Using the observed redshift and stellar mass, we calibrate the galaxy assembly parameter $B$ to bridge the initial and 
final stellar mass for each object (for more details, see \citealt{Inayoshi_2022a}).

Figure~\ref{fig:Msigma} presents the evolution tracks of the BHs and galaxies for the nine selected AGNs,
for each of which $\fstar=$ 0.05, 0.1 (solid curve), and 0.2 are considered, respectively.
We incorporate the individual evolution pathways at $z\lesssim 15$, drawing from tracks (see Figure~\ref{fig:Mevol}) with BH masses similar to the AGNs at their observed redshifts.
The population exhibiting a BH-to-stellar mass ratio near the local relation 
\citep[6 objects;][]{Ding_2023, Larson_2023, Harikane_2023} 
follows tracks that originate with small BH masses at $z\sim 15$ and subsequently approach the local relation. 
This implies that these BHs undergo moderate growth and maintain a relatively close proximity to 
the local relation throughout their evolution (e.g., \citealt{Habouzit_2022}; \citealt{Li_2022_HSC}).

In contrast, the remaining two objects follow distinct evolutionary trajectories. 
The BH in GS 3073 \citep{Ubler_2023} starts near the local relation and becomes overmassive 
by the observed epoch due to rapid mass accretion. 
The most massive BH at $z>6$, found in J0100+2802 \citep{Wu_2015}, has an evolutionary track that reaches the local relation 
at $z\simeq 8.6$ but then surpasses it by increasing the BH mass rapidly.
To understand the subsequent evolution of the distant BH at $z=8.68$ \citep{Larson_2023}, which might 
be a luminous quasar at $z\sim 6$, we extrapolate the BH and stellar masses and predict their future evolution.
We select two evolutionary tracks from Figure~\ref{fig:Mevol} that are closest to the BH of CEERS-1019
(note that the BH with $\mbh \simeq 10^6~\msun$ in GN-z11 recently reported by \citealt{Maiolino_2023} is also located close to
the two tracks).
By adopting these BH growth tracks and extrapolating the stellar masses to $z=6$, we find that this BH could either 
grow to be overmassive or maintain the BH-to-stellar mass ratio consistent with the local value
as indicated by the two dashed curves \citep{Agarwal_2013,Natarajan_2017,Inayoshi_2022a,Hu_2022b,Scoggins_2023}.

In addition, \cite{Bogdan_2023} reported X-ray detection in a high-$z$ galaxy (UHZ1), which has been confirmed as a $z=10.3$ galaxy with a stellar mass of 
$\sim 4-7\times 10^7~\msun$ \citep{Castellano_2023,Atek_2023}.
If the X-rays originate from an accreting BH and the luminosity is close to the Eddington value, the BH mass is inferred as $4 \times 10^7~\msun$.
Aware of the uncertainties, this object would challenge the BH seeding scenarios and their early growth.
The twelve AGNs discovered by \citet{Maiolino_2023Jades} are also overmassive relative to the host galaxies compared with the local relation.
We show an evolutionary track of the BH with the highest $\mbh/\mstar$ ratio among this sample. 
The pathway surpasses the local relation at $z \simeq 15$ and increases the BH mass to $\gtrsim 10^7~\msun$ by the observed epoch at $z\simeq 4.4$.
Further observational studies will improve our understanding on the initial conditions of BH-galaxy coevolution 
\citep[e.g.,][]{Inayoshi_2022a,Hu_2022b,Natarajan_2023,Pacucci_2023}, and the nature of the host galaxies that host overmassive BHs such as 
star formation efficiency and feedback processes \citep[e.g.,][]{Dekel_2023}.

\section{Summary}
\label{sec:summary}
In this paper, we expand upon the BH growth model developed in \cite{Li_2023}, linking the QLF at $z\simeq 6$ with those 
at lower redshifts ($z\simeq 5$ -- $4$).
This growth model captures the episodic nature of BH accretion and incorporates parameters that characterize
the duration of mass accretion, the Eddington ratio distribution function, and the mass dependency of BH accretion rates. 
Based on the previous work by \cite{Li_2023}, where the early assembly of seed and massive BHs is constrained in comparison with
the observed QLF \citep{Matsuoka_2018} and BHMF \citep{Willott_2010} at $z\simeq 6$,
we further extend the evolution of those BH populations down to $z= 5$ and $4$, and 
calibrate the growth model utilizing the QLF at each epoch \citep{Niida_2020,Akiyama_2018} across 
a wide UV magnitude range ($-29<\Muv<-24$).

Our best-fit parameters at each redshift interval vary substantially from those obtained at $z\gtrsim 6$, 
reflecting the different growth speeds and frequencies of accretion bursts in these stages. 
The rapid growth of massive BHs begins to decelerate at $5<z<6$ and is further stunted at $4<z<5$. 
This trend is indicated by both the characteristic Eddington ratio and the mass-dependent growth parameter
suppressing the growth of high-mass BHs.
We observe a saturation of BH mass growth at $\mbh \gtrsim 10^{10}~\msun$ at $z\lesssim 6$, consistent with 
the apparent maximum mass of observed SMBHs. 
We present the unobscured and total (unobscured + obscured) BHMF at $4<z<11$ according to our best-fit BH growth model. 
While our prediction overestimates the BH abundance at the lower mass range, it broadly agrees with current observational 
results in the high-mass end of the BHMF (see \citealt{He_2023}).
Our results offer a benchmark for future observational tests on the bulk shape of the BHMF at $z\gtrsim 5$, particularly their low-mass ends.
Moreover, we discuss the cosmic evolution of BH mass density by integrating the total BHMF, and find the result consistent 
with both X-ray observations at $z<5$ \citep{Ueda_2014} and the value inferred from recent observations of GN-z11 at $z=10.6$
\citep{Maiolino_2023}.

We construct evolutionary pathways for a large sample of BHs that grow from their initial seeding at $z\gtrsim 20$ to $z \simeq 4$, 
based on our best-fit model parameters. 
Our episodic BH growth model predicts both faint and bright phases for individual BHs throughout their evolution. 
This model successfully accounts for the observed quasar population at high redshifts, including the low-luminosity AGNs
recently detected in JWST observations \citep{Onoue_2023,Kocevski_2023,Ubler_2023,Harikane_2023,Larson_2023,Maiolino_2023}.

We further explore the early evolution of the BH-galaxy mass correlation, assuming that the stellar mass growth follows
that of the parent dark matter halo with a constant star formation efficiency.
Our results suggest two assembly pathways: (1) populations with a BH-to-stellar mass ratio near or below the local relation 
exhibit moderate BH growth before their observed epoch and remain close to the local relation, 
and (2) overmassive populations, in contrast, start with a mass ratio near the local relation and become overmassive
through rapid BH accretion at later epochs.

\acknowledgments
We acknowledge support from the National Natural Science Foundation of China (12073003, 11991052, 11721303, 12150410307, 11950410493), 
and the China Manned Space Project Nos. CMS-CSST-2021-A04 and CMS-CSST-2021-A06.
Y. M. was supported by the Japan Society for the Promotion of Science KAKENHI Grant Nos. 17H04830, 21H04494.

\appendix
\counterwithin*{equation}{section}
\renewcommand\theequation{\thesection\arabic{equation}}
\renewcommand{\thefigure}{A\arabic{figure}}
\setcounter{figure}{0}

\begin{figure*}
\begin{center}
\includegraphics[width=80mm]{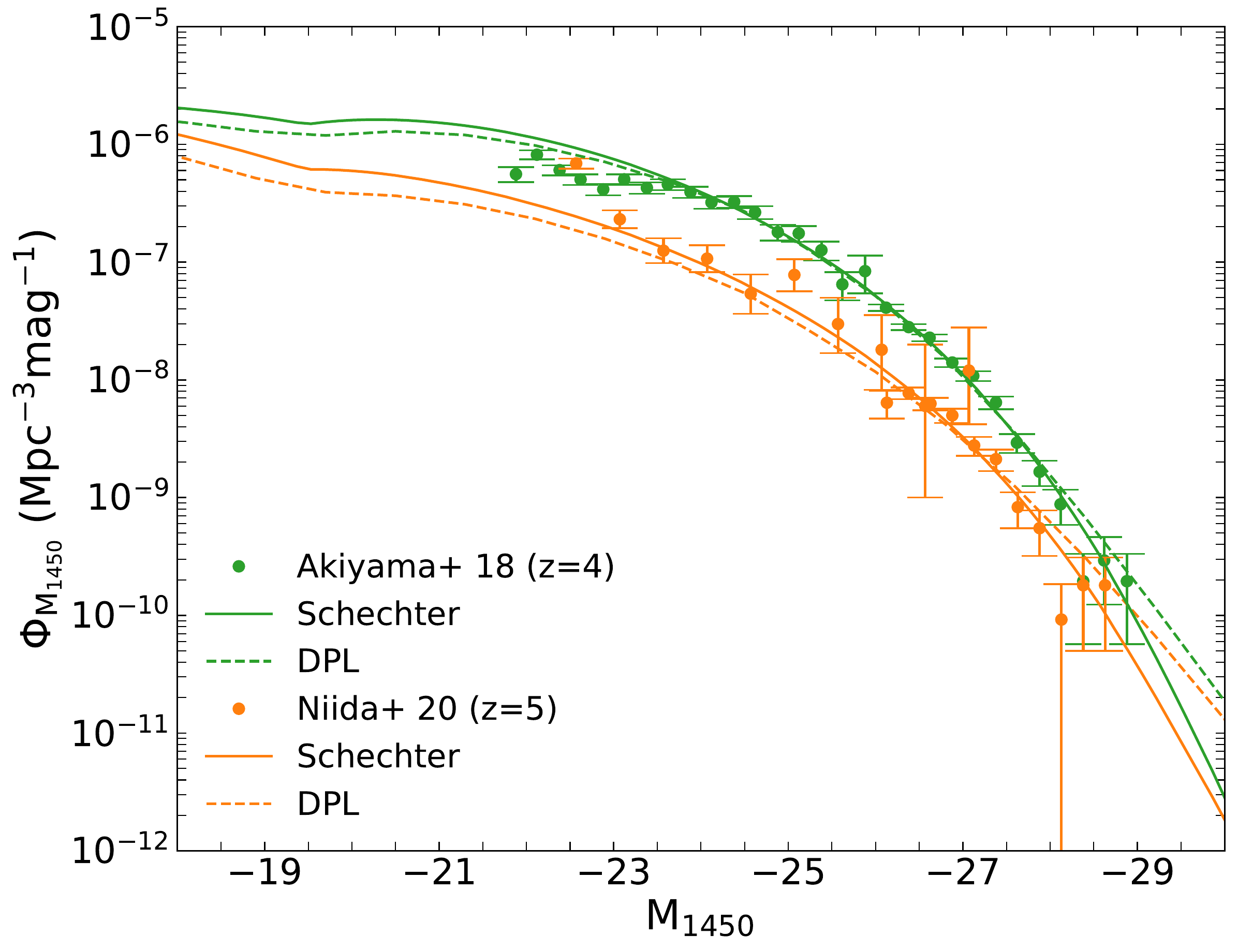}\hspace{5mm}
\includegraphics[width=83mm]{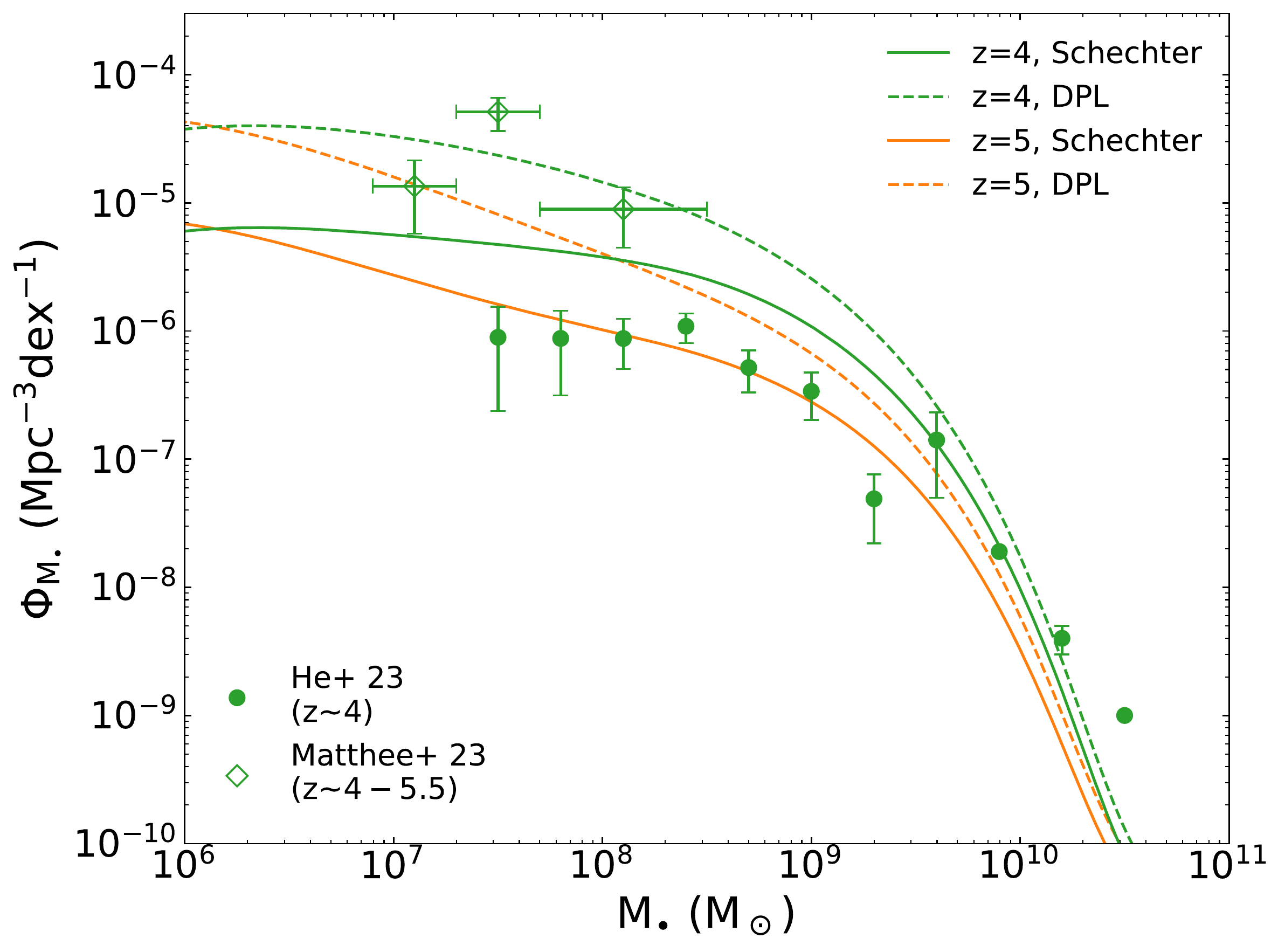}
\caption{{\it Left}: Unobscured quasar luminosity functions at $z=4$ (green) and 5 (orange) calculated from the best-fit model 
parameters with the input ERDF in a Schechter (solid) and DPL (dashed) shape, respectively.
The observed QLF data at the two redshifts are overlaid with the error bars \citep{Akiyama_2018,Niida_2020}.
{\it Right}: Unobscured BH mass functions at $z=4$ (green) and 5 (orange) for the Schechter (solid) and DPL (dashed) cases.
The observational results from \citet{He_2023} in the high-mass range and \citet{Matthee_2023} in the low-mass end are overlaid as filled and open symbols, respectively.}
\label{fig:DPL_SCH}
\end{center}
\vspace{4mm}
\end{figure*}

\begin{figure}
\begin{center}
\includegraphics[width=85mm]{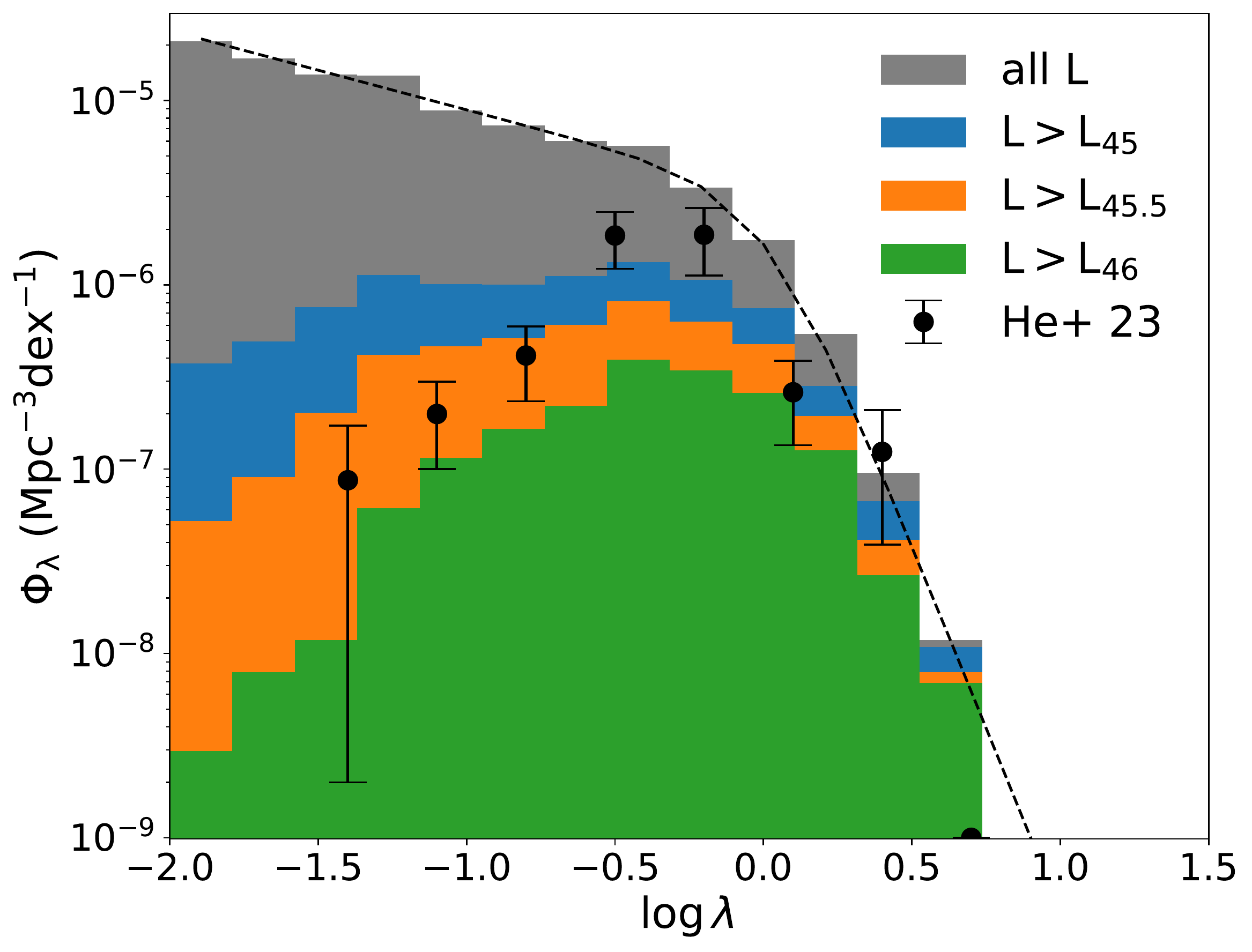}
\caption{Eddington ratio distribution functions at $z=4$ of the total BH population (grey, DPL) and selected by imposing different bolometric luminosity detection limits:
$L_\mathrm{bol} \geq L_{45}$ (blue), $L_{45.5}$ (orange), and $L_{46}$ (green).
The black symbols show the results from \citet{He_2023}, which is reconstructed from their $z\sim 4$ quasar sample with $L_\mathrm{bol} \gtrsim L_{45.5}$.}
\label{fig:lambdaDPL}
\end{center}
\vspace{4mm}
\end{figure}

\section{The ERDF in a double power-law}
The ERDF we adopt in the MCMC fitting in Section~\ref{sec:method} is in the form of Schechter shape characterized 
by two parameters: the slope of $\alpha$ in the low-$\lambda$ end and the characteristic Eddington ratio $\lambda_0$.
Here, we perform the MCMC fitting with a double-power-law (DPL) function shape for the ERDF
\beq
\label{eq:Pl_dpl}
\frac{\D P}{\D \ln \lambda} \propto
\left[\left(\frac{\lambda} {\lambda_l} \right)^{-(\alpha_l +1)} + \left(\frac{\lambda} {\lambda_l} \right)^{-(\beta_l +1)}\right] ^{-1},
\eeq
with three parameters (two slopes $\alpha_l$ and $\beta_l$, and the characteristic Eddington ratio $\lambda_l$).
Based on the $z=6$ BHMF taken from \cite{Li_2023}, we calculate the BHMFs and unobscured QLFs at $z=5$ and $4$,
and constrain the parameters of the DPL-shaped ERDF using the observational data at those redshifts.
In the Appendix, we set $\fseed=0.1$ and compare the best-fit result of the DPL case with the original Schechter case.

The left panel of Figure~\ref{fig:DPL_SCH} shows the best-fitted QLFs at $z=4$ (green) and $5$ (orange) for 
the cases with the different functional forms of the ERDF; DPL (dashed) and Schechter (solid) shapes.
Overall, both the results give similar fitting results and are consistent with the observed QLF data well.
While the DPL case suggests a higher abundance at the luminous end of the QLF, the difference is hardly distinguishable 
by the current observations.

The right panel of Figure~\ref{fig:DPL_SCH} presents the unobscured BHMFs at $z=4 $ and 5 calculated from the best-fit parameters
of the two ERDF forms, respectively.
The BH abundance in the DPL case tends to be higher than that in the Schecheter case because the best-fitted DPL form of the ERDF
allows a larger fraction of low-$\lambda$ populations compared to the Schechter case.
While the Schechter-like ERDF shows a better goodness of fit in comparison with the unobscured BHMF \citep{He_2023},
the DPL case would match the BHMF constructed with obscured AGNs reported with JWST observations \citep{Matthee_2023}.
The degeneracy in the ERDF form would be potentially solved by better determination of the BHMF at the low-mass end,
although AGN selection methods and unobscured/obscured AGN classification remain highly nontrivial and complex.
We will leave those issues for future investigation.

Finally, in Figure~\ref{fig:lambdaDPL} we demonstrate how the detection limit on the bolometric luminosity in quasar surveys affect the shape of the observed ERDF.
For the best-fit case with a DPL function form, as discussed in Section~\ref{sec:discussion_b},
we select quasars with $L_\mathrm{bol} \geq L_{45}=10^{45}$, $L_{45.5}=10^{45.5}$, and $L_{46}=10^{46}~\ergs$ at $z=4$.
Each histogram shows good agreement with the case with the Schechter-shaped ERDF (see Figure~\ref{fig:lambdaSch}).
Likewise, the DPL case with a threshold of $L_\mathrm{bol} \geq L_{45.5}$ matches the observational result in \citet{He_2023},
as in the Schechter case (see Section~\ref{sec:discussion_b}).


\newpage

\bibliography{main}{}
\bibliographystyle{aasjournal}


\end{document}